\documentclass[10pt]{iopart}
\usepackage{epsfig}

\def\modell{{\sl poly-gonato} model}
\def\lna{\langle\ln A\rangle}
\def\Xmax{$X_{max}$}
\def\gcm2{g/cm$^2$}
\def\knie{{\sl knee}}
\def\adhoc{{\sl ad-hoc }}

\def\Caption#1{\begin{minipage}[b]{0.49\textwidth}
                {\mathindent=10pt \caption{#1}}
	       \end{minipage}}

\def\ApP{{\it Astropart. Phys.\ }}      % Astroparticle Physics
\def\AA{{\it Astron. \& Astrophys.\ }}  % Astronomy and Astrophysics
\def\ApJ{{\it Ap. J.\ }}                % The Astrophysical Journal
\def\ICRC#1#2{{\it Proc. #1 Int. Cosmic Ray Conf. (#2)\ }}

\def\keyw#1{\vspace{10pt}
     \begin{indented}
     \item[]\rm Keywords: #1\par
     \end{indented}}

\def\remark#1{\vspace{10pt}
     \begin{indented}
     #1\par
     \end{indented}}

\begin{document}

\title[Total inelastic cross-sections and \Xmax\ in extensive air showers]
      {On total inelastic cross-sections and the average depth of 
       the maximum of extensive air showers}

\author{J\"org R. H\"orandel\footnote[3]{http://www-ik.fzk.de/$\sim$joerg}}

\address{University of Karlsruhe, Institut f\"ur Experimentelle Kernphysik,
         PO Box 3640, 76021~Karlsruhe, Germany}

%\remark{\item[] Submitted to \jpg }
\remark{\item[] Submitted to \jpg 24 Apr 2003, accepted 29 Aug 2003;\\
                Preprint astro-ph/0309010}
%\remark{\item[] in preparation for \jpg 22. April 2003}

\begin{abstract}
 The model dependence of the development of extensive air showers generated by
 high-energy cosmic-ray particles in the atmosphere is studied.  The increase
 of proton-proton and proton-air inelastic cross-sections and values for the
 elasticity are varied in the hadronic interaction model QGSJET.  Using the
 CORSIKA simulation program, the impact of these changes is investigated on air
 shower observables like the average depth of the shower maximum \Xmax\ and the
 number of muons and electrons at ground level.  Calculating the mean
 logarithmic mass from experimental \Xmax\ values, it is found that a moderate
 logarithmic increase of the proton-proton inelastic cross-section from
 $\sigma_{pp}^{inel}=51$~mb at $E_0=10^6$~GeV to $\sigma_{pp}^{inel}=64$~mb at
 $E_0=10^8$~GeV and an elasticity, additionally increased by 10\% to 15\%,
 describes the data best.  Using these parameters, the mean logarithmic mass
 $\lna$ derived from \Xmax\ measurements is compatible with the extrapolations
 of the results of direct measurements to high energies using the \modell.
\end{abstract}


\keyw{cosmic rays; air shower; hadronic cross-sections; 
      average depth of shower maximum; mass composition}

%\submitto{\jpg}

%\maketitle

\section{Introduction}
Cosmic rays above energies of $10^{14}$~eV are investigated by measurements of
extensive air showers in the atmosphere. These cascades of secondary particles
are generated by interactions of high-energy particles with air nuclei. While
the shower develops in the atmosphere the number of particles increases up to a
depth where the average energy of the secondaries equals a critical energy.
Beyond this point the number of particles decreases approximately
exponentially.  The depth where the cascade reaches the largest number of
charged particles is usually referred to as the depth of the shower maximum
\Xmax.

Two basic methods are used to examine air showers. In the first, the debris of
the particle cascade on the Earth's surface is observed in large detector
arrays, measuring electrons, muons, and hadrons, viz. particle distributions at
ground level. The second, calorimetric method investigates the longitudinal
development of the shower by registration of \v{C}erenkov or fluorescence
light, generated by the shower particles in the atmosphere, and eventually
\Xmax\ is derived.

Recently, the author has compiled the results of many air shower experiments
with the intention to estimate the best primary all-particle spectrum
(H\"orandel 2003). For this purpose, the energy spectra of individual elements
as directly measured at the top of the atmosphere below 1~PeV have been
extrapolated to high energy using power laws and assuming individual rigidity
dependent cut-offs.  The all-particle spectrum of these direct measurements was
compared with air shower observations at higher energies.  For the indirect
measurements it turned out that the individual energy scales had to be
renormalized to match the all-particle spectrum of direct measurements at
1~PeV.  Only small energy shifts were necessary, all within the errors
quoted by the experiments.  But most shifts had a negative sign, indicating an
overestimation of the primary energy. A likely explanation for this effect will
be discussed later in chapter~\ref{nenmusect}.

As a result all experiments yielded consistent all-particle energy spectra.
The extrapolations of the direct measurements have been fitted to the average
all-particle flux of the indirect measurements to determine the parameters of
the individual cut-offs mentioned above.  It has been shown, that the
experimental data can be parametrized consistently within a phenomenological
model, the \modell.  The mean logarithmic mass $\lna$ obtained from the model
is compatible with results from experiments measuring particle distributions at
ground level. But the mass composition disagrees with results from experiments
investigating the longitudinal shower development in the atmosphere.

A possible cause for the discrepancy might be, that the measured cascades
penetrate deeper into the atmosphere than predicted by the simulations, viz. the
codes predict a too small value of \Xmax.  A similar conclusion was drawn by
Erlykin and Wolfendale (2002).  
Investigating the muon production height with the KASCADE experiment, B\"uttner
\etal (2001) also find an indication for a deeper penetration of cascades into
the atmosphere as compared with predictions of the air shower simulation
program CORSIKA (Heck \etal 1998) with the high-energy interaction model QGSJET
(Kalmykov \etal 1997).

Reasons why the codes predict too small \Xmax -values may be numerous.
Objective of the present article is to study the influence of model parameters
like the inelastic cross-sections or the elasticity on the average value of
\Xmax\ and, consequently, on the mean logarithmic mass derived from it. Goal of
the investigations is to perhaps resolve the discrepancies in $\lna$ outlined
above.  

Some aspects of the impact of the complex system of parameters in simulation
codes on \Xmax\ are briefly summarized in chapter~\ref{litsect}.  Variations of
the interaction model QGSJET are described in chapter~\ref{crosssect}.  The
resulting changes of \Xmax\ and the consequences on $\lna$ are presented in
chapters~\ref{xmaxsect}  and \ref{lnasect}.  For completeness, the influence of
the changes on the number of electrons and muons at ground level is briefly
sketched in chapter~\ref{nenmusect}.

\section{Model parameters and the average depth of the shower maximum}
\label{litsect}
The longitudinal development of the nuclear cascade depends among others
essentially on following physics quantities: The inelastic cross-sections
$\sigma_{inel}$ of primary and secondary particles with air nuclei, the average
number of particles produced in an interaction --- the multiplicity $\mu$, and
the average fraction of energy transferred into secondary particles --- the
inelasticity $K$.  An increase of the inelastic cross-sections results in an
earlier development of the cascade. A similar effect has an increased
inelasticity. The particles lose more energy, hence the shower reaches its
maximum earlier in the atmosphere.  A reduction of the multiplicity produces
less particles in the first interactions. Due to energy conservation they will
be more energetic and the shower develops more slowly, i.e. the maximum is
deeper in the atmosphere.  Pajares \etal (2000) derived a parametrization for
the relative change of \Xmax\ as function of the changes of $\mu$ and $K$
\begin{equation}
 \frac{\Delta X_{max}}{X_{max}}\approx
      -\frac{1}{2}  \frac{\Delta\mu}{\mu} 
      -\frac{1}{10} \frac{\Delta K} {K}
 \quad.
 \label{xmaxmk}
\end{equation}
A similar relation for the dependence of \Xmax\ on inelastic cross-sections
will be given below in chapter~\ref{xmaxsect}.

A variety of models has been used to study the development of extensive air
showers and the effects of different parameters in the simulations on \Xmax\
have been elaborated in the literature. Some of the findings are summarized in
the following.

The average depths of the shower maximum obtained for different interaction
models implemented in CORSIKA --- i.e. DPMJET~2.5 (Ranft 1995 and 1999),
NEXUS~2 (Drescher \etal 2001), QGSJET~01, and SIBYLL~2.1 (Fletcher \etal 1994,
Engel \etal 1999) --- have been compared by Knapp \etal (2003). At \knie\
energies a spread of the models of about 50~\gcm2\ has been found for proton
induced showers and 20~\gcm2\ for primary iron nuclei. 

Similar results have been obtained by Fowler \etal (2001). The systematic
differences in \Xmax\ between different models in CORSIKA --- HDPM (Capdevielle
\etal 1992), QGSJET, SIBYLL, VENUS (Werner 1993) --- at 1~PeV are specified as
45~\gcm2\ for primary protons and 25~\gcm2\ for iron nuclei.

The longitudinal shower development for the hadronic interaction models SIBYLL
and QGSJET embedded in the framework of the MOCCA (Hillas 1997) and CORSIKA
programs has been explored by Pryke (2001). At 1~PeV the differences in \Xmax\
between CORSIKA/QGSJET and MOCCA with its internal hadronic interaction model
amount to 45~\gcm2\ for primary protons and to 30~\gcm2\ for iron induced
showers.  The differences between the interaction models are related to
distinct inelasticities: QGSJET produces more inelastic events, which lead to
less deeply penetrating showers. Using the MOCCA frame in connection with its
internal and the SIBYLL interaction model, the values differ only by about 5 to
10~\gcm2. This is compatible with estimates of Dickinson \etal (1999) who
obtained an increase of about 10~\gcm2\ from MOCCA/MOCCA to MOCCA/SIBYLL. 

Wibig (2001) discussed the influence of various hadronic interaction models ---
FRITIOF (Anderson \etal 1991), GMC (Wibig 1997), HDPM, QGSJET, SIBYLL, and
VENUS --- on the predictions for \Xmax\ using the shower simulation program
CORSIKA.  For a primary energy of 1~PeV he found differences in the order of
40~\gcm2\ for primary protons and 18~\gcm2\ for iron nuclei.  

A fast one-dimensional hybrid method has been used to simulate air showers by
Alvarez-Mu\~niz \etal (2002a). The models QGSJET~98 and two versions of SIBYLL
(1.7 and 2.1) have  been applied to describe the hadronic interactions.  For
SIBYLL~2.1 the inelastic proton-air cross-section rises faster with energy as
compared with QGSJET, whereas the inelasticities are almost equal in both
models.  On the other hand, the multiplicity of charged secondary particles
produced in proton-air collisions grows rapidly with energy for QGSJET.
Compared with QGSJET, for SIBYLL~2.1 the mean values of \Xmax\ of proton
induced showers are about 6~\gcm2\ larger at energies of 1~PeV and 10~\gcm2\ at
1~EeV. 

Gaisser \etal (1993) investigated the statistical model (Landau 1969, Fowler
1987), the Kopeliovich-Nikolaev-Potashnikova (KNP) model (Kopeliovich 1989),
and the mini-jet model (Gaisser and Halzen 1987).  The characteristic
differences between them are the energy dependences of the inelasticity. In the
statistical model the inelasticity decreases with energy, almost constant
values are obtained for the mini-jet model, and in the KNP model the
inelasticity grows with energy. At 1.25~EeV the depth of the maximum of proton
initiated showers varies by about 45~\gcm2. Due to the larger inelasticity the
showers develop faster in the KNP model as compared with the statistical model.
The latter represents the other extreme, generating relatively long showers.

The impact of internal parameters in the Quark Gluon String model (QGSJET) on
the average depth of the shower maximum was examined by Kalmykov \etal (1995).
Different implementations for the contributions of semihard processes lead to
differences in \Xmax\ for primary protons of about 18~\gcm2\ and 40~\gcm2\ at
1~PeV and 100~PeV, respectively.  The role of the production of charm particles
for the air shower development was found to be negligible, the estimated
differences are in the order of 3~\gcm2.

The effect of the multiplicity of secondary particles on \Xmax\ was scrutinized
by Anchordoqui \etal (1999) using the AIRES program (Sciutto 1998) with the
models SIBYLL and QGSJET. For both models the AIRES cross-sections have been
utilized in the calculations.  As a consequence of the lower inelasticity in
SIBYLL, the model produces fewer secondaries than QGSJET. Hence, there is a
delay in the shower development for  SIBYLL generated cascades. Primary protons
with energies of 1~PeV penetrate deeper into the atmosphere by about 45~\gcm2\
and 100~\gcm2\ at 1~EeV.

Capdevielle and Attallah (1995) pointed out how uncertainties in the
description of parton distribution functions of hadrons influence the
longitudinal development of air showers. The effects on inelastic
cross-sections, multiplicity and inelasticity were derived. The maximal
uncertainty in \Xmax\ was estimated to be at 1~EeV in the order of 95~\gcm2\
for proton induced showers and 40~\gcm2\ for iron primaries.

A second step cascading mechanism was discussed in the geometrical multichain
model by Wibig (1999). Second step cascading is defined as the interaction of a
wounded nucleon of one nucleus with another nucleon from the same nucleus
before hadronization occurs. These internuclear cascades result in an increase
of the hadron air inelasticity for heavy nuclei. Results of CORSIKA simulations
with this interaction model indicate that this mechanism results in an earlier
shower development for iron nuclei of 10~\gcm2\ to 30~\gcm2\ at energies of
10~PeV to 10~EeV.

New effects in high-energy interactions like percolation, quark gluon plasma,
or string fusion were anticipated by Pajares \etal (2000).  The authors
concluded that these effects dump the multiplicity and increase the
inelasticity. This leads to larger values of \Xmax.

Erlykin and Wolfendale (2002) linked the observed discrepancies in $\lna$ to
two effects in nucleus-nucleus collisions not been taken into account before: A
few percent energy transfer into the electromagnetic component due to
electron-positron pair production or electromagnetic radiation of quark gluon
plasma and a small slow-down of the cascading process in its initial stages
associated with the extended lifetime of excited nuclear fragments. The latter
displaces the shower deeper into the atmosphere.

To summarize, several hypothetical effects have been described which could
change the longitudinal shower development and their influence is not easy to
segregate from each other. The maximum model ambiguities, taking into account
all considerations as discussed, amount in the \knie\ region to about $\Delta
X_{max}\approx45$~\gcm2\ for primary protons and 25~\gcm2\ for iron induced
showers. These uncertainties increase with energy to $\Delta
X_{max}\approx100$~\gcm2\ for proton induced showers at 1~EeV.  In principle,
the quoted ambiguities could be reduced (by some 30\%) as some model approaches
can be rejected by theoretical arguments and/or comparison with accelerator or
cosmic-ray data.  In addition to the model ambiguities discussed so far, there
are experimental uncertainties, which will be discussed in
chapter~\ref{xmaxsect}.

\section{Inelastic hadronic cross-sections and other model parameters} 
   \label{crosssect}
Among the various parameters which control the shower development, the
inelastic cross-sections and the elasticity will be scrutinized more closely.
The actual situation can be recapitulated as follows.  

\subsection{Inelastic cross-sections}
Inelastic cross-sections for proton-proton interactions are of great interest
for particle physics. They have been measured in detail in collider
experiments, but studied also at highest energies using cosmic rays.  At
energies exceeding $\sqrt{s}=1$~TeV the knowledge about the increase of the
proton-proton cross-section as function of energy is limited by experimental
errors of both, collider and cosmic-ray experiments.  The insufficient
knowledge of the high-energy cross-sections restricts the reliability of
simulation programs to calculate the interactions of high-energy cosmic rays.

A recent review of $\bar{p} p$ cross-sections from collider experiments has
been given by Hagiwara \etal (2002).  At present, the highest energies at
colliders are available at the Tevatron ($\sqrt{s}=1.8$~TeV).  There, values
for the total $\bar{p} p$ cross-section disagree by about 10\% between 
different experiments, i.e.  
$71.71\pm2.02$~mb  (E-811, Avila \etal 1999), 
$72.8 \pm 3.1$~mb  (E-710, Amos \etal 1992), and
$80.03\pm 2.24$~mb (CDF,   Abe \etal 1994a).

In cosmic-ray experiments the attenuation of the proton flux entering the
atmosphere is used to derive the inelastic proton-air cross-section.  Then, the
latter is utilized to derive the proton-proton cross-section using the Glauber
theory (Glauber and Matthiae 1970).  In air shower experiments three basic
methods are applied to obtain cross-sections: \\
1) The ratio of the proton flux at the top of the atmosphere $\Phi_0$ to the
flux of surviving protons $\Phi_g$ measured below the atmosphere at a depth
$x$.  From these quantities the interaction lengths of protons in air
$\lambda_{p-air}(E)$ is determined using the relation $ \Phi_g(E,x)/\Phi_0(E) =
\exp[-x/\lambda_{p-air}(E)]$. This method needs high fluxes of penetrating
primary protons and, therefore, can be applied at relatively low energies only,
see Yodh \etal (1983).\\
The second and third method take advantage of the fact that cosmic-ray
protons interact in the atmosphere at rates which decrease exponentially with
increasing depth. \\
2) For a fixed primary energy and zenith angle the distribution of the average
depth of the shower maximum has an exponential tail with a slope given by the
attenuation length $\Lambda$, see e.g. (Baltrusaitis \etal 1984) or (Gaisser
\etal 1993).\\
3) The last method to derive $\Lambda$ utilizes the zenith angle distributions
of the shower intensity for fixed primary energies, as described by Hara \etal
(1983).

The attenuation length is connected with the interaction length $\lambda$ by
the relation
\begin{equation}
 \Lambda(E) = k(E) \lambda_{p-air}(E) 
         = k(E) \frac{\langle A_{air}\rangle~m_p}{\sigma_{p-air}^{inel}(E)}  
	 \quad.
 \label{sigfun}	   
\end{equation}
$k(E)$ is a model dependent proportional factor, which among others depends on
the mean inelasticity of the interactions $K(E)$.  $\sigma_{p-air}^{inel}(E)$
is the total inelastic proton-air cross-section, $\langle A_{air}\rangle=14.5$
the effective atomic weight of air with the proton mass $m_p$.

The conversion of the attenuation lengths to cross-sections depends essentially
on theoretical assumptions on $k(E)$, for a discussion and theoretical
motivation see e.g. (Kopeliovich \etal 1989), (Engel \etal 1998), or (Block
\etal 2000).  For example, using a different value for $k$ in equation
\eref{sigfun}, total inelastic cross-sections obtained by Baltrusaitis \etal
(1984) and Honda \etal (1993) around c.m. energies $\sqrt{s}\approx10$~TeV have
been reduced by 10\% to 15\% by Block \etal\hspace{-1.5mm}.  The original value
of the Fly's Eye experiment
$\sigma_{p-air}^{inel}=530\pm66$~mb at $\sqrt{s}=30$~TeV
has been rescaled to
$\sigma_{p-air}^{inel}=460\pm40$~mb.

The subsequent conversion of proton-air to proton-proton cross-sections depends
on theoretical uncertainties as well. The details required to apply Glauber
theory in this context have been discussed by Gaisser \etal (1987) as well as
by Wibig and Sobczy\'nska (1998).  The latter apply a geometrical scaling
hypothesis, use an exact Glauber formalism and conclude, that the proton-proton
cross-sections reported by Baltrusaitis \etal and Honda \etal are overestimated
by about 10\%.

Besides the theoretical uncertainties, all methods require that the primary
cosmic-ray flux contains a sufficient fraction of protons, e.g. Hara \etal
(1983) derive cross-sections under the assumption of a contribution of at
least 10\% protons.  Thus, additional systematic errors have to be taken into
account concerning the unknown mass composition at high energies.

\begin{figure}[bt] \centering
 \epsfig{file=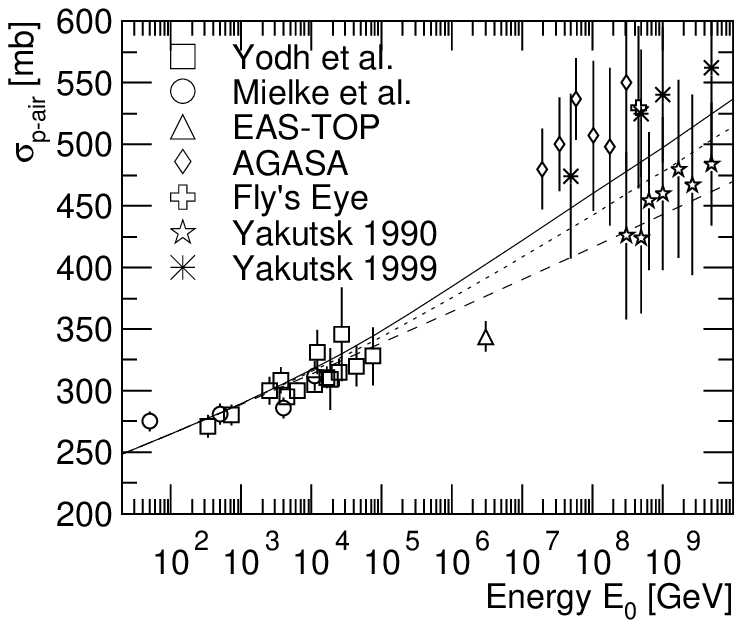,width=0.5\textwidth}
 \Caption{Inelastic proton-air cross-sections versus primary energy.
	  For references of the experimental values see text.  The lines
	  represent calculated cross-sections for three different versions of
	  the model QGSJET~01, model~1 (original QGSJET) (\full),
	  model~2~(\dotted), and model~3 (\dashed), see text.
          \label{pair} }
\end{figure} 

In early cosmic-ray work an increase of the inelastic cross-section as function
of energy has been found by Yodh \etal (1972, 1973).  In the energy range up to
100~TeV the measured flux of unaccompanied hadrons at ground level has been
used to calculate lower bounds for the proton-air cross-sections by Yodh \etal
(1983) combining several experimental results and also by Mielke \etal (1994)
using the KASCADE prototype calorimeter.  Particle numbers measured with the
EAS-TOP experiment have been used to deduce a value for the inelastic
proton-air cross-section at \knie\ energies (Aglietta \etal 1997).  At the
highest energies $\sigma_{p-air}^{inel}$ has been derived from the shower
attenuation length $\Lambda$ by the Fly's Eye experiment (Baltrusaitis \etal
1984) using the exponential tail of \Xmax\ distributions and by the AGASA
experiment (Honda \etal 1993) measuring particle numbers at ground level.
Dyakonov \etal (1990) derived lower limits for cross-sections from
investigations of the Yakutsk data, applying two methods, fits to the
exponential tail of \Xmax\ and observing the zenith angle distribution of the
shower intensity for fixed energies.  A second analysis of the Yakutsk data
(Knurenko \etal 1999) takes into account the tail of the \Xmax\ distributions
at fixed energies.  The results obtained are shown in \fref{pair} as function
of laboratory energy $E_0$.  

\begin{figure}[bt]\centering
 \epsfig{file=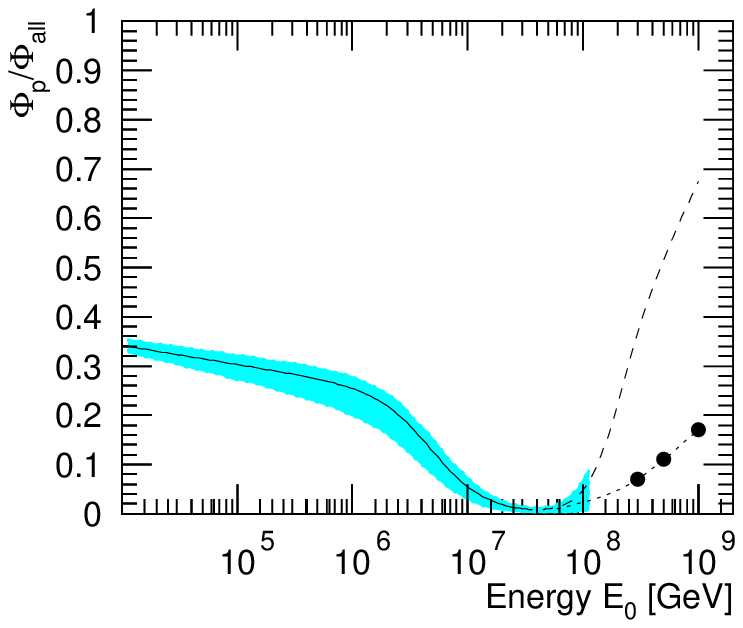, width=0.5\textwidth}
 \Caption{Relative fraction of protons in cosmic rays versus primary energy
 according to the \modell\ (\full). The band illustrates the expected error. 
 The line above $3\cdot10^7$~GeV is an upper limit assuming the \adhoc
 component being protons only (\dashed).  The points are obtained by fits to
 measured \Xmax\ distributions of the Fly's Eye experiment, see
 section~\ref{xmaxdistsect}.  They are connected by a line to guide the eye
 (\dotted).
          \label{pfract} }
\end{figure}

The experimental values shown in \fref{pair} have been derived assuming a
sufficiently large contribution of protons in primary cosmic rays. The validity
of this assumption shall be examined next.  For this reason, the fraction of
protons as function of energy is shown in \fref{pfract}.  It has been obtained
from the analysis of direct measurements and air shower data applying the
\modell, as outlined in the introduction.  The solid line represents the ratio
$\Phi_p/\Phi_{all}$ for the cosmic rays which shall be denoted, with
precaution, as the "galactic component". The error band indicates the
uncertainties expected from the model. Since the average all-particle spectrum
is reasonably well known, the error indicated depends essentially on the
parametrization of the proton flux and its errors.  According to the \modell\
the galactic component is characterized by cut-offs of the flux of individual
elements at energies proportional to their nuclear charge
($\hat{E}_Z=\hat{E}_p\cdot Z$), starting with protons at an energy
$\hat{E}_p=4.5$~PeV. Above $10^8$~GeV this galactic component is not sufficient
to describe the observed all-particle spectrum within the model.  Hence, an
additional component has to be introduced {\sl ad hoc} in order to account for
the measured flux values.  Above $3\cdot 10^7$~GeV the dashed line gives an
upper limit for the proton fraction, assuming the \adhoc component  being
protons only. The proton fraction decreases with energy from 35\% at $10^4$~GeV
to below 1\% at $3\cdot10^7$~GeV. Beyond this energy the upper limit increases
again. However, taking the error band into account, at least between $10^7$ and
$10^8$~GeV protons contribute only little to the all-particle flux ($<5\%$).
Thus, one of the premises to derive proton-air cross-sections in this energy
region is weakened and, maybe, the values obtained from air shower measurements
have to be corrected.

In order to check the impact of inelastic cross-sections on the development of
extensive air showers, the cross-sections are altered within a particular
model.  The present studies have been carried out using the simulation program
CORSIKA version 6.0190. Investigations by the KASCADE group (Antoni \etal 1999,
Milke \etal 2001) indicate that the interaction model QGSJET~98 describes the
measured data most consistently.  Hence, the latest version of this code, i.e.
QGSJET~01, has been chosen for the simulations. Within this program the
parametrizations for the inelastic hadronic cross-sections have been modified
in order to change their energy dependence at high energies, the contribution
of mini-jets has been reduced. The changes affect all inelastic cross-sections,
like proton-proton, proton-air, and pion-air at the same time.  In addition, a
parameter had to be readjusted which essentially influences the multiplicity of
produced particles in order to match pseudo-rapidity distributions of collider
experiments. This implies a reduction of the average number of charged
particles produced.  

In the following the original QGSJET code is referred to as model~1, while the
two modifications with smaller cross-sections are labeled model~2 and 3,
respectively. In addition, for model~3 the average transverse momentum has
been reduced and the elasticity has been increased, these changes are combined
in model~3a.

\begin{figure}[bt]\centering
 \epsfig{file=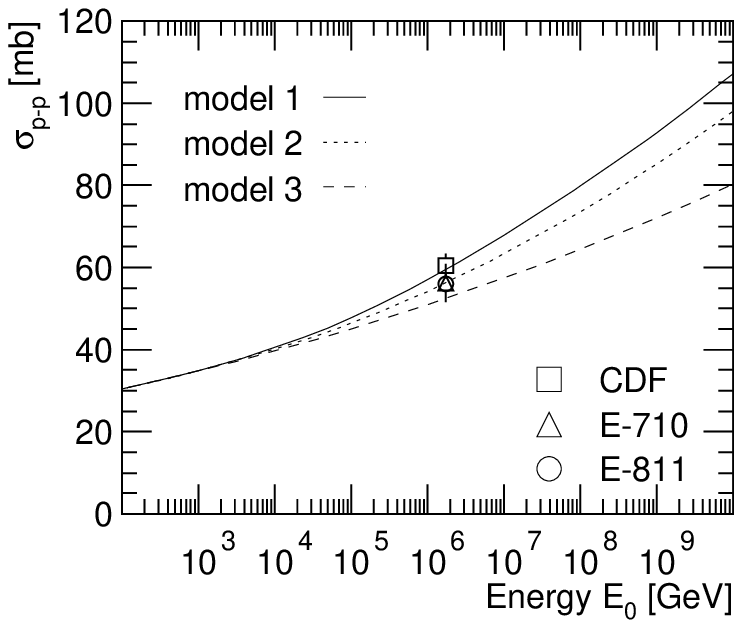,width=0.5\textwidth}
 \Caption{Inelastic proton-proton cross-section versus laboratory energy
	  for three versions of the model QGSJET~01. The data points are for
	  $\bar{p} p$ interactions from three experiments at the Tevatron, see
	  text.
          \label{pp} }
\end{figure}

The inelastic cross-sections for models~1 to 3 for proton-proton collisions are
shown in \fref{pp} as function of laboratory energy.  At $10^8$~GeV the
cross-sections vary between
$\sigma_{pp}^{inel}=80$~mb for the original QGSJET model and
$\sigma_{pp}^{inel}=64$~mb for model~3.
At the Tevatron ($E_0=1.7\cdot10^6$~GeV) for the total elastic $\bar{p} p$
cross-section the values
$\sigma_{\bar{p} p}^{el}=15.79\pm0.87$~mb (E-811, Avila \etal 1999),
$\sigma_{\bar{p} p}^{el}=16.6 \pm1.6 $~mb (E-710, Amos  \etal 1990), and
$\sigma_{\bar{p} p}^{el}=19.70\pm0.85$~mb (CDF,   Abe   \etal 1994b)
have been measured. Together with the total cross-sections mentioned above,
values for the total inelastic cross-section are obtained as shown in the
figure. Model~1 corresponds to the CDF value, model~2 to the results of E-710
and E-810, while model~3 marks approximately the $1 \sigma$ lower error bound
for the last two experiments.

Next, the model predictions are compared with air shower measurements.  The
proton-air cross-sections for models~1 to 3 have been plotted in \fref{pair}
and are compared with experimental results. At $10^8$~GeV values between
$\sigma_{p-air}^{inel}=460$~mb for the original QGSJET~01 (model~1) and
$\sigma_{p-air}^{inel}=416$~mb for model~3 are obtained.  At low energies up to
$10^5$~GeV the results by Yodh \etal (1983) and Mielke \etal (1994) are
compatible with the QGSJET cross-sections for all three cases considered.  The
cross-section obtained by the EAS-TOP collaboration is still lower, even when
compared with model~3.  On the other hand, the results from AGASA and Fly's Eye
are about 1 to 2 $\sigma$ above the calculated curve for model~3.  Taking into
account the low proton content of primary cosmic rays in this energy region
according to the \modell, the experimental values are likely too large.  The
results of the EAS-TOP and AGASA experiments rely on the measurements of
particle numbers at ground level.  As has been pointed out by Alvarez-Mu\~niz
\etal (2002a), intrinsic shower fluctuations strongly influence the
cross-sections obtained from such measurements.  The experimental uncertainties
may be illustrated by two analyses of the Yakutsk data. The cross-sections
obtained by Dyakonov \etal (1990) are found to be compatible with model~3,
while on the other hand, the more recent values of Knurenko \etal (1999) are
about $1\sigma$ above model~3.  Summarizing, no serious disagreement between
experimental values and the cross-sections for model~3 can be stated.  

\begin{figure}[bt] \centering
 \epsfig{file=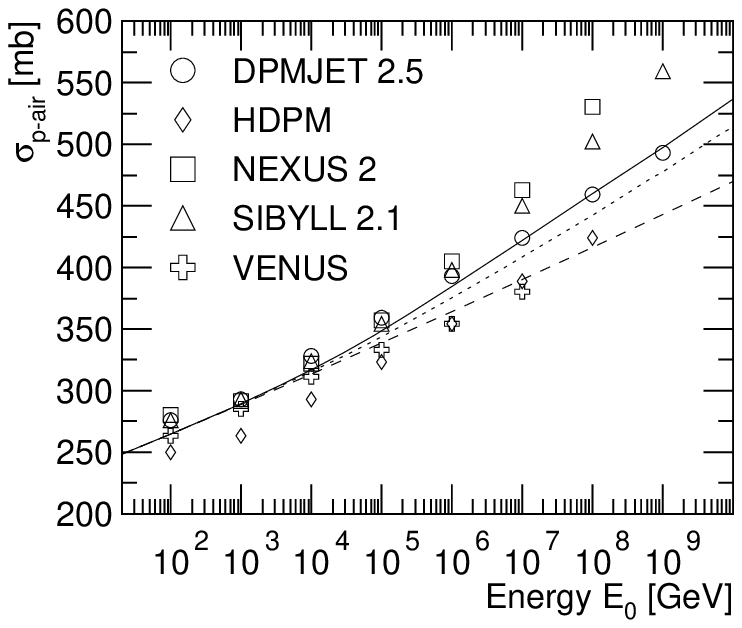,width=0.49\textwidth}
 \epsfig{file=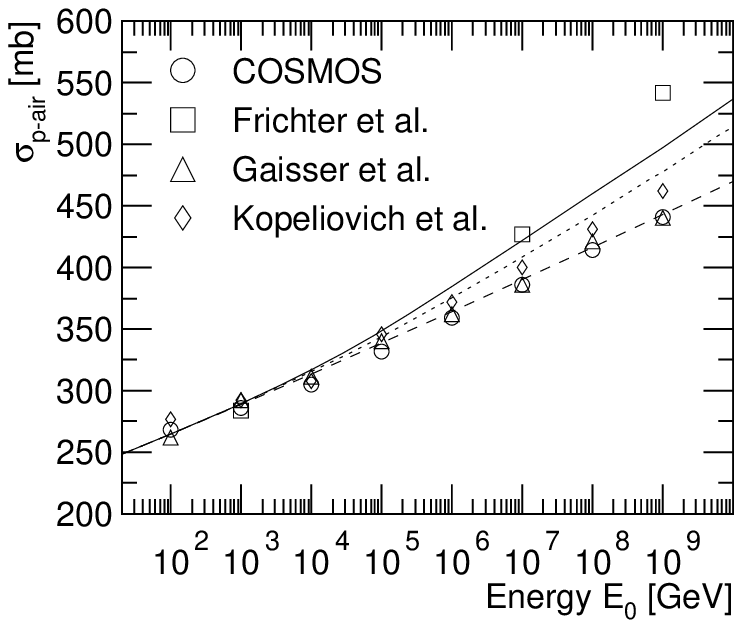,width=0.49\textwidth}
 \caption{Inelastic proton-air cross-sections versus primary energy.
	  The lines represent calculated cross-sections for three different
	  versions of the model QGSJET~01, model~1 (original QGSJET) (\full),
	  model~2~(\dotted), and model~3 (\dashed).  For comparison,
	  cross-sections of interaction models implemented in CORSIKA
	  (left-hand graph) and other models (right-hand graph) are shown, for
	  references see text.
	 }
 \label{pairm}         
\end{figure} 

In \fref{pairm} the cross-sections for models~1 to 3 are compared with results
of other model calculations. In the left-hand panel the values for the models
DPMJET~2.5, HDPM, SIBYLL~2.1, VENUS (Knapp \etal 2003), and NEXUS~2 (Bossard
\etal 2001) as implemented in CORSIKA are shown. Compared with model~3 the
cross-sections for DPMJET, NEXUS, and SIBYLL grow faster as function of energy.
At energies below $10^7$~GeV the HDPM cross-sections are smaller than the
values for model~3, while VENUS exhibits nearly the same behavior as model~3.

The right-hand graph summarizes further theoretical considerations.
Kopeliovich \etal (1989) extrapolate results from collider experiments and
calculate inelastic proton-air cross-sections. Their results for a QCD pomeron
with an asymptotic intercept $\Delta=0.097$ are presented, these cross-sections
are slightly larger than those for model~3.  Also Frichter \etal (1997)
extrapolate results from accelerator measurements to high energies. They use
relative low inelasticities and in turn obtain cross-sections larger as the
previous model. But they admit that such low inelasticities are unable to
account for the Fly's Eye data reported by Gaisser \etal (1993) and Bird \etal
(1993).  Similar values as for model~3 are obtained by Huang \etal (2003) using
the COSMOS simulation code.  Gaisser \etal (1993) investigate the falling slope
of the observed \Xmax\ distribution as measured by the Fly's Eye experiment and
conclude, that this decrement is consistent with a near linear $\lg(s)$ energy
dependence of the inelastic proton-air cross-section. Their results are shown
in the figure as well and are found to coincide with the values for model~3.
The authors conclude further, when the extrapolations of proton-proton
cross-sections from collider experiments to high energies by Block \etal (1992)
are converted to proton-air cross-sections (Gaisser \etal 1987), one obtains
values which are slightly above the $\lg(s)$ dependence mentioned above.  Also
these values are compatible with those of model~3.

\begin{table}[bt] \centering 
 \caption{Inelastic proton-proton and proton-air cross-sections as function of 
	  laboratory energy for three versions of the high-energy hadronic
	  interaction model QGSJET~01.  Model~1 corresponds to the original
	  QGSJET, in model~2 and 3 the total inelastic cross-sections have been
	  reduced, see text.\\}
 \label{sigtab}
 \small
 \begin{tabular}{r rrcrrcrr} \hline
         & \multicolumn{8}{c}{Total inelastic cross-sections [mb]} \\
  \multicolumn{1}{c}{Energy} &
    \multicolumn{2}{c}{model 1} &&
    \multicolumn{2}{c}{model 2} &&
    \multicolumn{2}{c}{model 3} \\ \cline{2-3} \cline{5-6} \cline{8-9}
  $E_0$ [GeV] & 
    $\sigma_{pp}^{inel}$ & $\sigma_{p-air}^{inel}$ &&
    $\sigma_{pp}^{inel}$ & $\sigma_{p-air}^{inel}$ &&
    $\sigma_{pp}^{inel}$ & $\sigma_{p-air}^{inel}$ \\ \hline
 % wqtab.tex erzeugt mit paw/wq/writewqtab.kumac
$10^2$     &   30 &  265 &&  30 &  265 &&  30 &  265 \\
$10^3$     &   35 &  289 &&  35 &  289 &&  35 &  289 \\
$10^4$     &   40 &  317 &&  40 &  315 &&  40 &  314 \\
$10^5$     &   48 &  349 &&  46 &  344 &&  45 &  339 \\
$10^6$     &   57 &  385 &&  54 &  375 &&  51 &  364 \\
$10^7$     &   68 &  422 &&  63 &  408 &&  57 &  390 \\
$10^8$     &   80 &  460 &&  74 &  443 &&  64 &  416 \\
$10^9$     &   93 &  498 &&  85 &  478 &&  72 &  443 \\
$10^{10}$  &  107 &  536 &&  98 &  514 &&  80 &  470 \\
 \hline 
 \end{tabular} 
\end{table}

The inelastic cross-sections for both, proton-proton and proton-air
interactions according to models~1 to 3 are listed in \tref{sigtab} as
function of energy for reference.

\subsection{Other model parameters}

\begin{figure}[b] \centering
 \epsfig{file=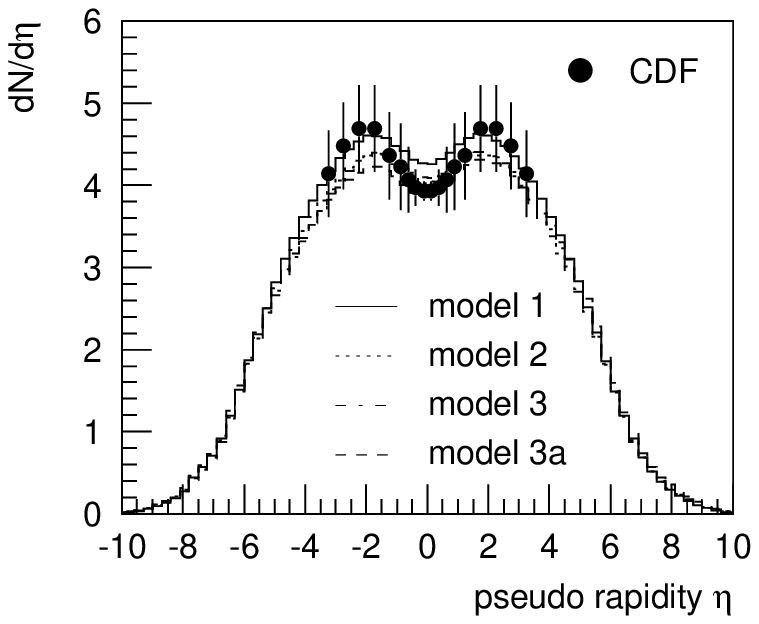,width=0.49\textwidth}
 \Caption{Pseudo-rapidity distribution of charged particles in
	  $p$-$\bar{p}$ collisions at $\sqrt{s}=1.8$~TeV for four versions
	  of QGSJET compared with Tevatron data (CDF, Abe \etal 1990).
          \label{prapidity}}         
\end{figure} 

The reduction of the contribution of mini-jets also influences the
pseudo-rapidity~$\eta$.  Therefore, another parameter --- mainly reducing the
multiplicity --- had to be readjusted in order to match measurements at the
Tevatron ($\sqrt{s}=1.8$~TeV ), shown in \fref{prapidity} together with the
predictions of the models~1 to 3a.  For central collisions ($\eta\approx0$) the
values increase from model~3a to model~1. While for large $|\eta|$ values the
distributions for all models almost coincide.  In conclusion, all four models
yield very similar distributions, which describe well the CDF-measurements (Abe
\etal 1990), taking their error bars into account.

\begin{figure}[bt] \centering
 \epsfig{file=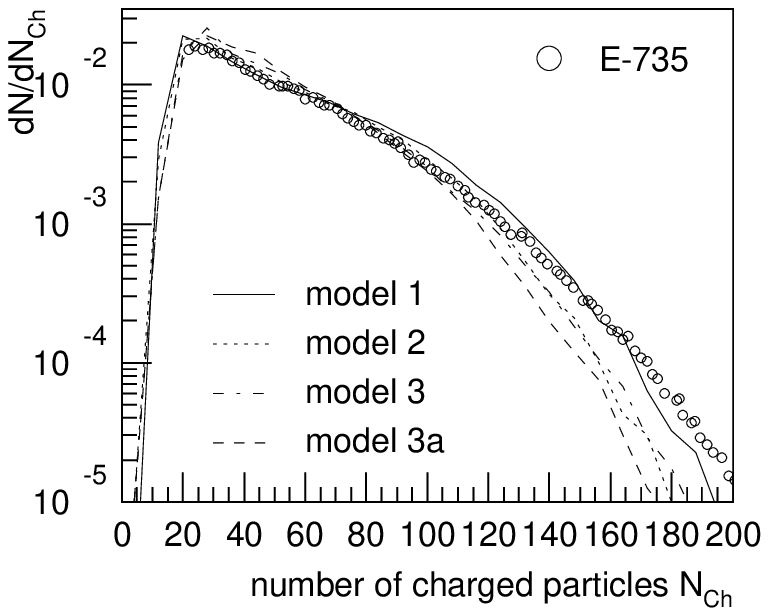,width=0.49\textwidth}
 \epsfig{file=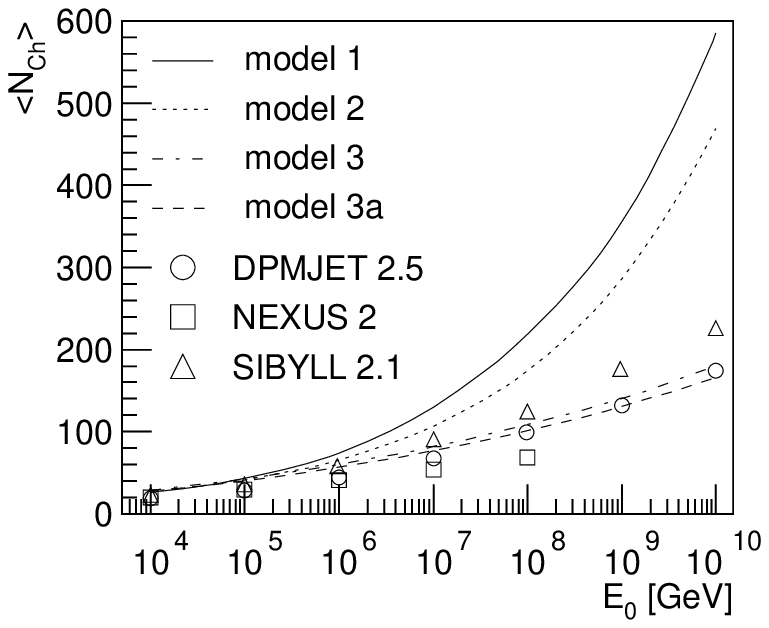,width=0.49\textwidth}
 \caption{Number of charged particles produced in $p$-$\bar{p}$ collisions
	  at $\sqrt{s}=1.8$~TeV for the QGSJET modifications together with
	  measurements of E-735 (Alexopoulos \etal 1998) (left-hand side).
	  Average number of particles produced in $\pi$-$^{14}$N collisions as
	  function of laboratory energy for four versions of the model QGSJET
	  and values for three other models implemented in CORSIKA (Knapp \etal
	  2003) (right-hand graph).
	 }
 \label{ncharged}         
\end{figure} 

The frequency distributions of the number of charged particles simulated for
proton-antiproton collisions at $\sqrt{s}=1.8$~TeV are plotted on the left-hand
side of \fref{ncharged}. The results obtained with the four models are very
similar.  A closer look reveals, the right-hand tail of the distributions is
slightly steeper for models~3 and 3a as compared with model~1. At the same
time, the average multiplicity is reduced by up to 5\% from $49.4\pm0.1$
(model~1) to $47.1\pm0.1$ for model~3 and $46.9\pm0.1$ for model~3a.  The model
predictions are compared with measurements at the Tevatron (E-735, Alexopoulos
\etal 1998). The data have been normalized to the calculated values for model~1
over a range of multiplicities starting at the peak of the distributions around
$N_{Ch}\approx20$. At medium multiplicities ($\approx 100$) model~1 slightly
overestimates the production probability, while at high multiplicities all four
models are below the measured values.

Of great importance for the air shower development are pion-nitrogen
collisions. The average multiplicity of charged particles $\langle
N_{Ch}\rangle$ for these interactions is shown in \fref{ncharged} (right-hand
side) as function of laboratory energy.  For models~3 and 3a the average
multiplicity is significantly lower as compared with model~1. The differences
amount to more than a factor of three at the highest energies shown. Compared
with other interaction models like DPMJET, NEXUS, or SIBYLL, relatively large
values for $\langle N_{Ch}\rangle$ are obtained with the original QGSJET at
high energies (Knapp \etal 2003).  On the other hand, the multiplicities
obtained with models~3 and 3a coincide with the values for DPMJET~2.5.

\begin{figure}[bt] \centering
 \epsfig{file=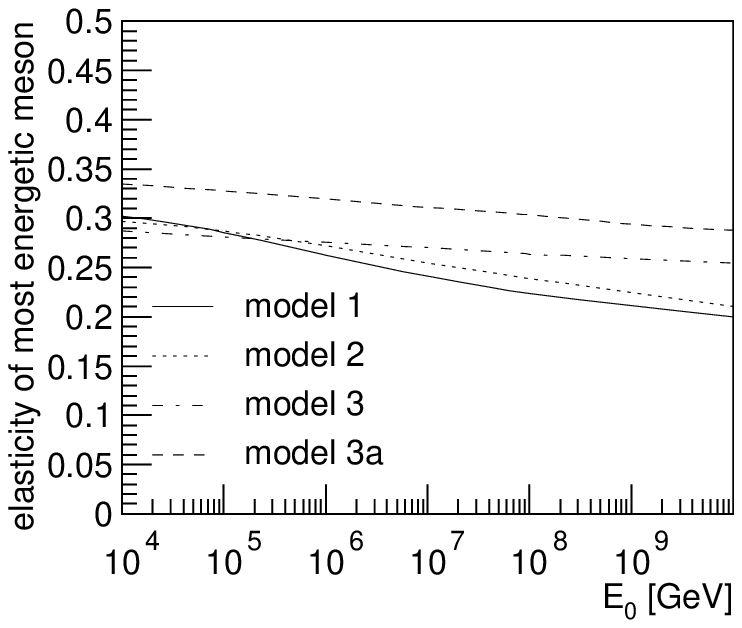,width=0.49\textwidth}
 \Caption{Average elasticity of most energetic meson in $\pi$-$^{14}$N 
          interactions versus laboratory energy for four versions of the model
	  QGSJET.
          \label{elasticity}}         
\end{figure} 

The average elasticity of the most energetic meson in pion-nitrogen
interactions is plotted in \fref{elasticity} as function of laboratory energy
for the four models. As can be seen in the figure, the reduction of the
mini-jet contribution, as described above, also increases the elasticity at
high energies. In model 3a the elasticity has been increased additionally by
10\% to 15\% relative to model 3 to check the influence of this parameter on
the shower development.

\begin{figure}[b] \centering
 \epsfig{file=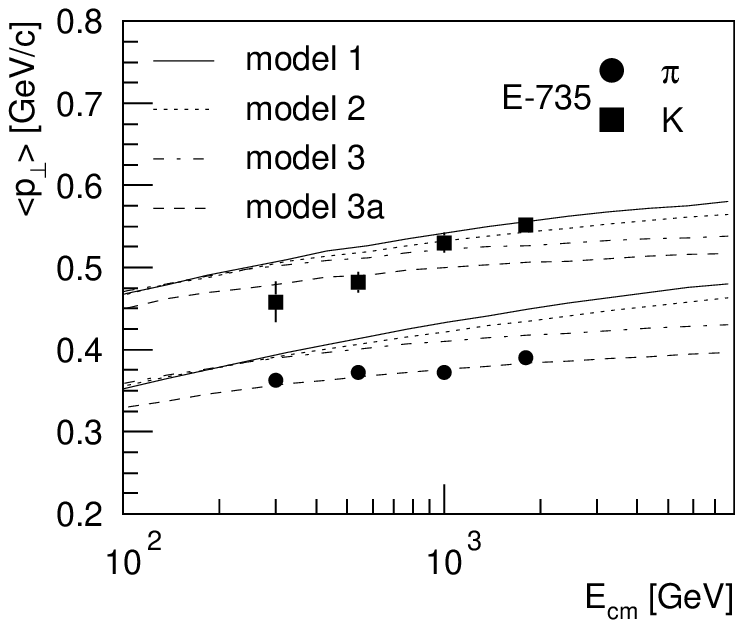,width=0.49\textwidth}
 \epsfig{file=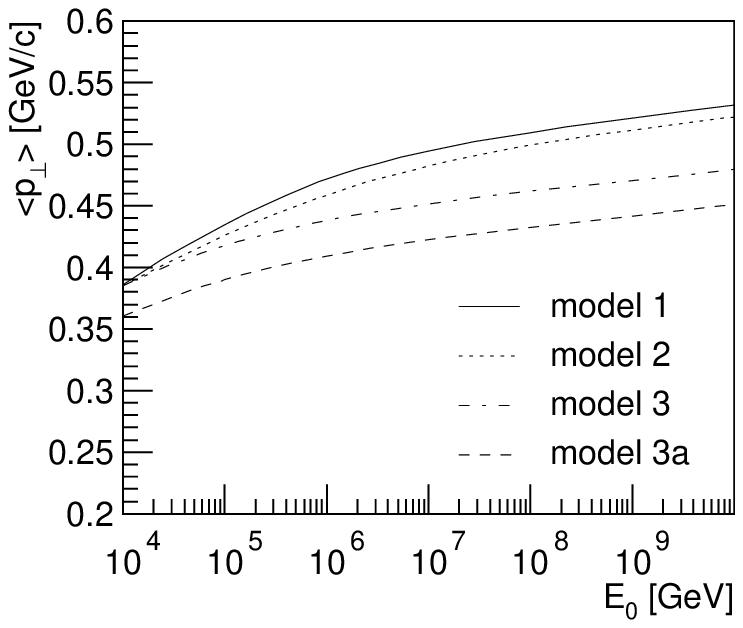,width=0.49\textwidth}
 \caption{Left-hand panel: Average transverse momentum of pions (bottom) and 
	  kaons (top) in $p$-$\bar{p}$ collisions as function of c.m. energy
	  for four versions of the model QGSJET compared with Tevatron data
	  (E-735, Alexopoulos \etal 1993).
	  Right-hand panel: Average transverse momentum of charged particles
	  in $\pi$-$^{14}$N interactions as function of laboratory energy.
	 }
 \label{pt}         
\end{figure} 
Reducing the inelastic cross-sections in QGSJET via a reduction of the
mini-jet contribution also reduces the average transverse momentum $\langle
p_\bot\rangle$. Average values of $p_\bot$ for charged pions and kaons in
proton-antiproton collisions are shown in \fref{pt} (left-hand graph) as
calculated with the different versions of QGSJET in the energy region of the
Tevatron. Measurements from experiment E-735 are depicted for comparison. 
The average
$p_\bot$ obtained with models~1 to 3 is up to 15\% larger than the measurements
for pions.  Therefore, the average $p_\bot$ predicted with model~3a has been
adjusted to the experimental values shown.  Unfortunately, the trend of the
energy dependence of $p_\bot$ for kaons is less well described by all models
discussed. But, since pions are clearly dominant for the air shower
development, emphasis is given to a correct description of the transverse
momenta for pions.

The average transverse momentum of charged particles in pion-nitrogen
collisions is shown in \fref{pt} (right-hand graph) as function of laboratory
energy.  At the highest energies the values for model~3a are about 15\% below
the average $p_\bot$ as calculated with model~1.

In summary, it may be concluded that the four models exhibit distinct
differences in certain quantities. Individual experimental results are
described better by one or another model. But in a general view, all four
versions of QGSJET are compatible with results from collider and air shower
experiments, or at least no serious disagreement between the models and the
measurements can be stated.

\section{The average depth of the shower maximum} \label{xmaxsect}
In this chapter, the consequences of changing the inelastic cross-sections and
the other modified parameters on the longitudinal development of air showers
shall be investigated. For this purpose, in the energy range from $10^5$~GeV up
to $3\cdot 10^{10}$~GeV, 500 showers induced by primary protons and 100 by iron
nuclei have been simulated for each energy in steps of half a decade using the
CORSIKA program. To describe the high-energy hadronic interactions the
models~1 to 3a have been used together with the GHEISHA code for energies below
80~GeV. The discussion of the results is divided in two sections, on average
\Xmax\ values and on \Xmax\ distributions.

\subsection{Average \Xmax-values}

\begin{figure}[b]\centering
 \epsfig{file=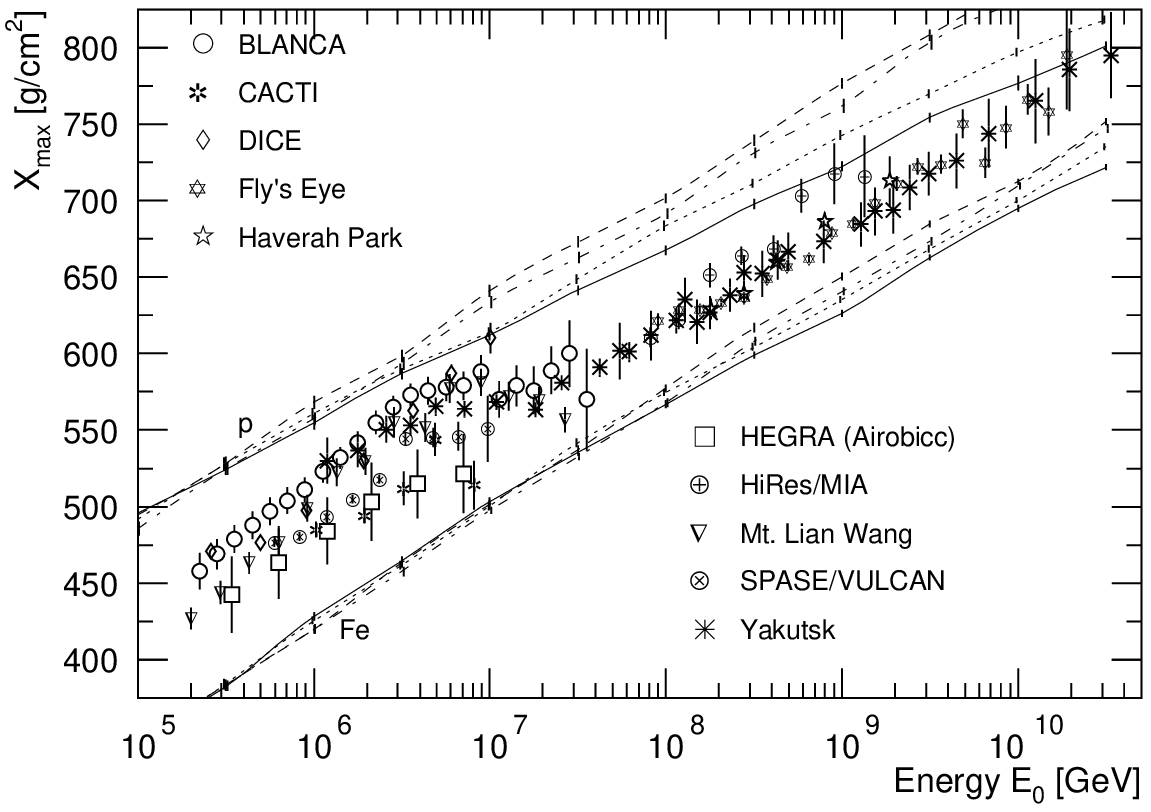, width=0.8\textwidth}
 \caption{Average depth of the shower maximum versus primary energy.
          Results are shown from the experiments
	  BLANCA (Fowler \etal 2001),
	  CACTI (Paling \etal 1997),
	  DICE (Swordy \etal 2000),
	  Fly's Eye (Bird \etal 1994),
	  Haverah Park (Watson 2000),
	  HEGRA (Arqueros \etal 2000),
	  HiRes/MIA (Abu-Zayyad \etal 2000),
	  Mt. Lian Wang (Cha \etal 2001),
	  SPASE/VULCAN (Dickinson \etal 1999), and
	  Yakutsk (Dyakonov \etal 1993, Knurenko \etal 2001).
	  The data are compared with results from CORSIKA simulations for
	  primary protons and iron nuclei using different versions of QGSJET:
	  model~1 (original QGSJET) (\full), model~2 (\dotted), model~3
	  (\chain), and model~3a (\dashed).
	 }
 \label{xmax}         
\end{figure}

The average depth of the shower maximum as obtained by the calculations is
shown in \fref{xmax} as function of energy for primary protons and iron nuclei
for the four models considered.  The original QGSJET yields the smallest \Xmax\
values, with model~3a the showers penetrate deepest into the atmosphere.  The
values change at $10^8$~GeV from
$X_{max}=668\pm3$~\gcm2\ to
$X_{max}=702\pm3$~\gcm2\ for proton induced showers and from
$X_{max}=567\pm2$~\gcm2\ to                
$X_{max}=577\pm2$~\gcm2\ for iron primaries.
\Xmax\ is listed in \tref{xmaxtab} as function of energy for the four versions
of the QGSJET model.

The simulated values are compared with experimental results from various
experiments in the figure.  Two different techniques are used to measure \Xmax,
namely the imaging technique, using telescopes to obtain a direct image of the
shower and the non-imaging method, in which the height of the shower maximum is
derived from the lateral distribution of the \v{C}erenkov light (or electrons
in case of Haverah Park) measured at ground level. The DICE, Fly's Eye, and
HiRes experiments use the imaging method, while all other experiments belong to
the second group.  The data show systematic differences of $\approx 30$~\gcm2\
at 1~PeV increasing to $\approx65$~\gcm2\ close to 10~PeV. A more detailed
discussion of the measurements is given elsewhere (H\"orandel 2003).  Some of
the experimental uncertainties may be caused by changing atmospheric
conditions. The experiments measure a geometrical height which has to be
converted into an atmospheric depth. Measuring longitudinal atmospheric
profiles during different seasons, Keilhauer \etal (2003) found that the
atmospheric overburden for a fixed geometrical height (e.g. 8~km a.s.l.) varies
by at least 25~\gcm2.  

\begin{table}[bt] \centering
 \caption{Average depth of the shower maximum for primary protons and iron 
	  nuclei calculated with CORSIKA, using different versions of the
	  interaction model QGSJET, see text.\\}
 \label{xmaxtab}
 \small
 \begin{tabular}{r 
                 r@{$\pm$}r r@{$\pm$}r c 
		 r@{$\pm$}r r@{$\pm$}r c
		 r@{$\pm$}r r@{$\pm$}r c
		 r@{$\pm$}r r@{$\pm$}r   } \hline
	       & \multicolumn{14}{c}{Average depth of the shower maximum 
	                             \Xmax~[\gcm2]} \\	 
  Energy       & \multicolumn{4}{c}{model 1} &
               & \multicolumn{4}{c}{model 2} &
               & \multicolumn{4}{c}{model 3} &
               & \multicolumn{4}{c}{model 3a} \\ 
	             \cline{2-5} \cline{7-10} \cline{12-15} \cline{17-19}
  $E_0$  [GeV] & \multicolumn{2}{c}{p} & \multicolumn{2}{c}{Fe} &
               & \multicolumn{2}{c}{p} & \multicolumn{2}{c}{Fe} &
               & \multicolumn{2}{c}{p} & \multicolumn{2}{c}{Fe} &
               & \multicolumn{2}{c}{p} & \multicolumn{2}{c}{Fe} \\ \hline
%%%%%%%%%%%%%%%%%%%%%%%%%%%%%%%%%%%%%%%%%%%%%%%%%%%%%%%%%%%%%%%%%%%%%%%%
% xmax.tex erzeugt mit xmaxtab.kumac und script formatxmax in paw/knee 
%%%%%%%%%%%%%%%%%%%%%%%%%%%%%%%%%%%%%%%%%%%%%%%%%%%%%%%%%%%%%%%%%%%%%%%% 
$1.00\cdot10^5$& 496&   6 & 349&   3 && 492&   4 & 346&   3 && 486&   4 & 338&   4 && 494&   5 & 345&   3 \\ 
$3.16\cdot10^5$& 524&   4 & 382&   3 && 528&   4 & 384&   3 && 525&   4 & 383&   3 && 528&   4 & 384&   3 \\ 
$1.00\cdot10^6$& 555&   5 & 428&   3 && 561&   4 & 426&   3 && 558&   4 & 420&   3 && 568&   4 & 420&   3 \\ 
$3.16\cdot10^6$& 587&   4 & 464&   3 && 590&   3 & 462&   3 && 594&   4 & 458&   3 && 599&   4 & 463&   3 \\ 
$1.00\cdot10^7$& 612&   3 & 503&   3 && 614&   3 & 500&   3 && 634&   4 & 498&   2 && 641&   4 & 501&   3 \\ 
$3.16\cdot10^7$& 641&   3 & 535&   2 && 649&   3 & 542&   3 && 663&   4 & 533&   2 && 673&   4 & 537&   3 \\ 
$1.00\cdot10^8$& 668&   3 & 567&   2 && 684&   3 & 575&   2 && 692&   3 & 568&   3 && 702&   3 & 577&   2 \\ 
$3.16\cdot10^8$& 698&   3 & 599&   2 && 711&   3 & 604&   3 && 731&   4 & 608&   3 && 739&   4 & 617&   3 \\ 
$1.00\cdot10^9$& 722&   3 & 626&   2 && 743&   3 & 635&   2 && 762&   4 & 640&   3 && 776&   3 & 650&   2 \\ 
$3.16\cdot10^9$& 754&   3 & 663&   2 && 770&   3 & 668&   3 && 804&   5 & 673&   3 && 808&   3 & 684&   2 \\ 
$1.00\cdot10^{10}$& 777&   5 & 695&   2 && 797&   3 & 700&   2 && 830&   3 & 710&   2 && 842&   3 & 712&   2 \\ 
$3.16\cdot10^{10}$& 801&   3 & 722&   2 && 818&   3 & 735&   2 && 866&   4 & 747&   3 && 870&   3 & 752&   2 \\ 
%%%%%%%%%%%%%%%%%%%%%%%%%%%%%%%%%%%%%%%%%%%%%%%%%%%%%%%%%%%%%%%%%%%%%%%%
 \hline
 \end{tabular}
\end{table}

The ambiguities between the different experiments are of the same order as the
systematic differences between individual models as discussed in
chapter~\ref{litsect}, where ambiguities of about 45~\gcm2\ for proton and
about 25~\gcm2\ for iron induced showers have shown up.  It can be noticed that
up to 4~PeV the measured \Xmax\ values increase faster as function of energy
than the simulated values for the original QGSJET (model~1). With the
modifications in model~3a the simulated showers penetrate deeper into the
atmosphere and the increase is almost parallel to the data.

An estimate for the dependence of $\Delta X_{max}/X_{max}$ on the multiplicity
and the inelasticity has been given in equation \eref{xmaxmk}.  The present
investigations confirm the dependence of \Xmax\ on the inelasticity.
The correlations between changes in the inelastic proton-proton as well as
proton-air cross-sections and \Xmax\ can be estimated by
\begin{equation}
 \frac{\Delta X_{max}}{X_{max}} \approx
       -\frac{2}{7}\, \frac{\Delta\sigma_{pp}^{inel}   }{\sigma_{pp}^{inel}   }
       \quad\mbox{and}\quad
 \frac{\Delta X_{max}}{X_{max}} \approx
       -\frac{5}{7}\, \frac{\Delta\sigma_{p-air}^{inel}}{\sigma_{p-air}^{inel}}
 \quad.
\end{equation}
Since the changes of the parameters in QGSJET affect all cross-sections
simultaneously, the dependencies on one particular cross-section can not be
disentangled.

In the superposition model of air showers, the development of a cascade induced
by a primary particle with energy $E$ and nuclear mass number $A$ is described
by $A$ proton induced subshowers of energy $E/A$. Hence, the shower maximum of
a proton induced shower $X_{max}^p$ should be equal to that for an iron primary
$X_{max}^{Fe}$ at higher energy
\begin{equation}\label{superposfun}
 X_{max}^p(E)=X_{max}^{Fe}(E\cdot A_{Fe}) \quad.
\end{equation}
The quantity $\Delta(E)$ defined as
\begin{equation}\label{dfun}
 \Delta(E)\equiv X_{max}^{Fe}(E\cdot A_{Fe})-X_{max}^{Fe}(E)
          = X_{max}^p(E)-X_{max}^{Fe}(E)
\end{equation}
specifies the increase of the \Xmax\ curve in the energy interval
$(E,A_{Fe}\cdot E)$, it relates to the frequently used elongation rate ${\cal
D}\equiv dX_{max}/d\lg E$ as $\Delta(E)={\cal D}(E) \lg(A_{Fe})$.  The
right-hand side of equation \eref{dfun} follows immediately from equation
\eref{superposfun}. This indicates, that the slopes of the \Xmax\ curves are
correlated with the difference between the proton and iron curves at a given
energy.  Such a behavior is visible in \fref{xmax}.  For model~3a the curves
for protons and iron nuclei are almost straight lines in the logarithmic plot.
In the energy range shown, $\Delta(E)$ changes from about 150~\gcm2\ to
approximately 120~\gcm2\ only.  On the other hand, the logarithmic slope of the
two curves for model~1 decreases significantly with rising energy and the
difference between protons and iron diminishes from 
$\Delta(3\cdot10^5~\mbox{GeV})\approx150$~\gcm2\ to
$\Delta(3\cdot10^{10}~\mbox{GeV})\approx80$~\gcm2.    

\begin{figure}[bt]\centering
 \epsfig{file=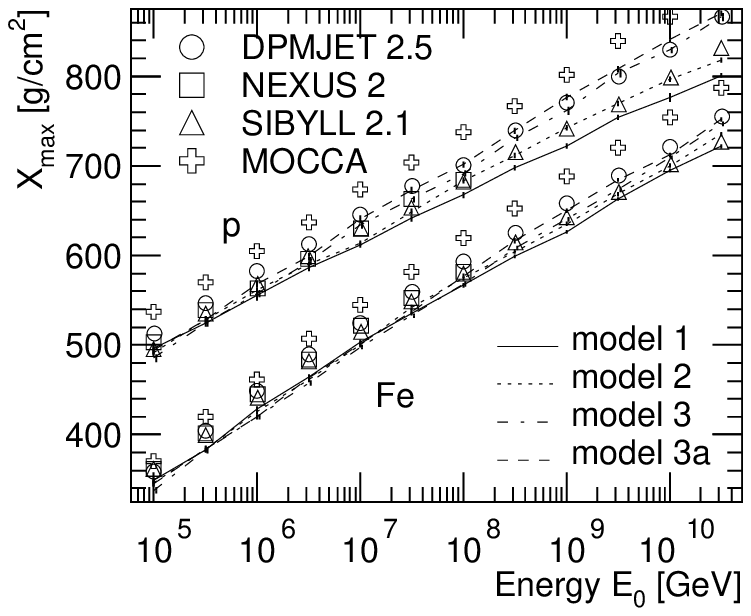,width=0.5\textwidth}
 \Caption{Average depth of shower maximum versus primary energy.
	  Predictions of simulations for proton and iron induced showers using
	  CORSIKA with the interaction models DPMJET, NEXUS, and SIBYLL (Knapp
	  \etal 2003) as well as the MOCCA program (Pryke 2001) are compared
	  with results of the modifications of QGSJET~01 (models~1 to 3a).
          \label{xmaxmod} }
\end{figure}

In the previous chapter cross-sections for several models have been presented.
It is interesting to compare the depths of the shower maxima calculated with
these codes to the depths obtained with the QGSJET modifications.  \Xmax\
values for primary protons and iron nuclei for the interaction models
DPMJET~2.5, NEXUS~2, and SIBYLL~2.1 have been calculated by Knapp \etal (2003)
using the CORSIKA program. The results are compared in \fref{xmaxmod} to values
for models~1 to 3a. SIBYLL predicts relatively early developing cascades with
values between models~1 and 2. On the other hand, DPMJET generated showers
penetrate deeper into the atmosphere similar to model~3a.  The validity of the
NEXUS code is guaranteed by its authors up to $10^8$~GeV and, correspondingly,
\Xmax\ up to this energy is plotted. The depths calculated with this code are
very similar to the SIBYLL predictions.  Depths simulated with the MOCCA code
using the internal interaction model (Pryke 2001) are significantly larger as
compared with the other codes.  With the original QGSJET~01 (model~1) the
shortest development of the cascades is obtained.

In \fref{pairm} it has been shown, that the inelastic cross-sections for SIBYLL
rise faster as compared with DPMJET.  This is compatible with deeper
penetrating showers for DPMJET, as can be inferred from \fref{xmaxmod}.  How
also the other parameters, like the multiplicity, effect the longitudinal
development can be seen as an example with DPMJET. Its showers
penetrate deeper into the atmosphere than those generated with model~1, despite
the almost identical cross-sections (see \fref{pairm}). However, the
multiplicity of charged particles grows faster with energy for QGSJET as
compared with DPMJET, e.g. at $10^9$~GeV the multiplicities differ by about a
factor of 3 (see \fref{ncharged}).

\subsection{\Xmax-distributions} \label{xmaxdistsect}
\begin{figure}[bt]\centering
 \epsfig{file=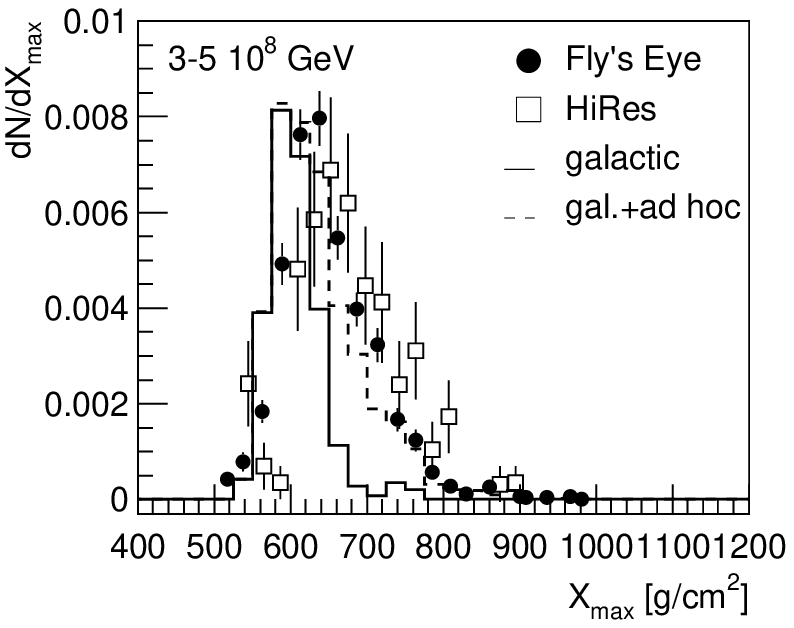,width=0.49\textwidth}
 \epsfig{file=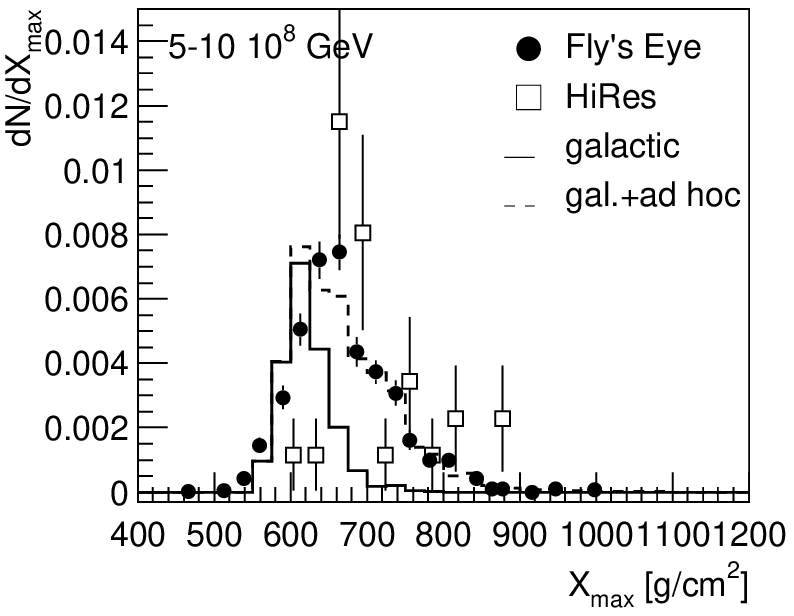,width=0.49\textwidth}
 \epsfig{file=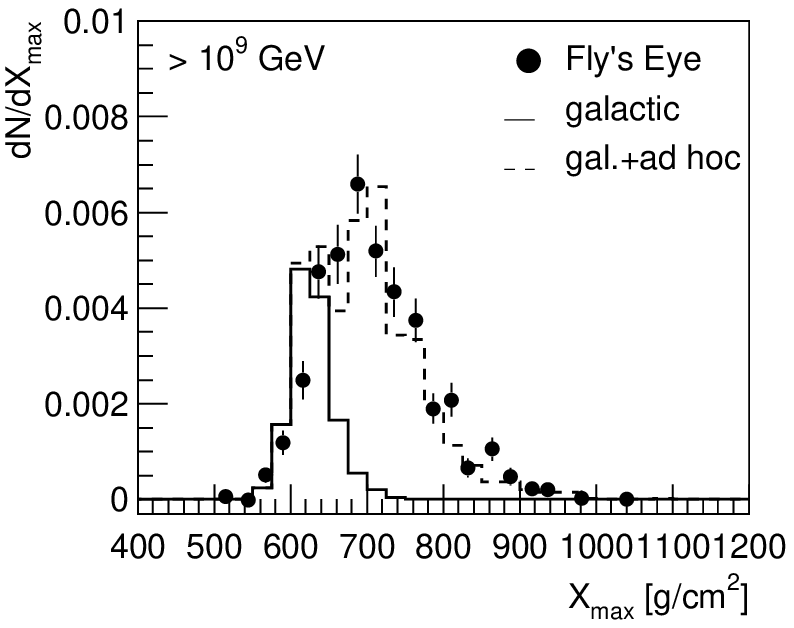,width=0.49\textwidth}
 \Caption{Distribution of the depth of the shower maximum \Xmax\ measured by 
	  the Fly's Eye (Gaisser \etal 1993) and HiRes (Abu-Zayyad \etal 2001)
	  experiments.  The measured values are compared with simulated results
	  using model~3a.  Galactic component according to the \modell\ (\full)
	  and galactic plus \adhoc component (\dashed), see text.
          \label{xmaxdist}  }
\end{figure}

In addition to the average depths, probability distributions of \Xmax\ values
have been published and can be investigated.  In the energy range from
$3\cdot10^8$ to beyond $10^9$~GeV experimental \Xmax-distributions from the
Fly's Eye (Gaisser \etal 1993) and HiRes (Abu-Zayyad \etal 2001)  experiments
are presented in \fref{xmaxdist}.  Early \Xmax\ distributions of the Yakutsk
experiment (Dyakonov \etal 1990) are not compatible with more recent average
\Xmax\ values (Dyakonov \etal 1993), shown in \fref{xmax}, and are therefore
not taken into account.

The results of the measurements are compared with calculated \Xmax\
distributions using model~3a for primary protons, helium, oxygen, iron, and
ultra-heavy nuclei.  The CORSIKA code allows primary nuclei with mass numbers
$A\le56$. The \Xmax\ distributions for ultra-heavy nuclei have been estimated
assuming $X_{max}\propto\ln A$, based on the simulated values for protons and
iron nuclei.  For each energy interval the mass composition is taken according
to the \modell, i.e. the abundances are taken as predicted and listed in
\tref{fracttab}.  Distributions for five mass groups, represented by the
nuclei mentioned above, have been calculated.  The resulting
\Xmax-distributions are shown in \fref{xmaxdist} as solid histograms. For
the three energy intervals the left-hand side of the Fly's Eye distributions is
described reasonably well by the galactic component.  It should be stressed
that the agreement has not been achieved by a fit to the distributions.
Instead, the mass composition as obtained with the \modell\ has been chosen.  

When comparing the same Fly's Eye data to results obtained with the original
QGSJET code, an offset $\Delta X_{max}=30$~\gcm2\ had to be added to the
simulated \Xmax\ values in order to reconcile the measured with the simulated
distributions (H\"orandel 2003).  A shift in the opposite direction has been
performed by Gaisser \etal (1993) in order to achieve agreement between
calculations with the KNP model and the Fly's Eye data, the simulated values
have been shifted by $\Delta X_{max}=-25$~\gcm2.  As can bee seen in
\fref{xmaxdist}, with model~3a no artificial offset is necessary to describe
the Fly's Eye distributions at the side of low \Xmax.  This indicates
experimental evidence for deeper penetrating showers as obtained with model~3a.

The galactic component accounts only for a fraction of the observed cosmic rays
in the energy range considered in \fref{xmaxdist}.  As mentioned in
chapter~\ref{crosssect}, close to $10^8$~GeV the heaviest elements of the
galactic component reach their cut-off energies and a new component starts to
dominate the all-particle spectrum.  For the additional component only the flux
but not the mass composition can be predicted.  Assuming only protons for this
component cannot explain the experimental \Xmax\ distributions. Hence, helium
and the CNO group have been added. A fit to the Fly's Eye data using the
simulated distributions for protons, helium, and oxygen yields the dashed
histograms as the sum of the galactic and the \adhoc component.  The relative
abundances obtained are listed in \tref{fracttab} in brackets.  As can be seen
in the figure, the dashed histograms reproduce well the Fly's Eye
distributions, indicating a mixed mass composition for the \adhoc component.
The contribution of the CNO group seems to be essential to describe the peak
region of the experimental distributions.  Protons account for the long,
asymmetric tail towards large values of \Xmax.  The steep gradients on the
left-hand side originate in the relative narrow distributions for the ultra
heavy elements of the galactic component with their small intrinsic
fluctuations in the shower development.

\begin{table}[bt]\centering
 \caption{Relative abundances $[\%]$ for groups of nuclei with charge number $Z$
   for different primary energies $E_0$ for the galactic component according to
   the \modell.  Values in brackets are for the \adhoc component in order
   to explain \Xmax\ measurements of the Fly's Eye experiment.\\}
 \label{fracttab}
 \small
 \begin{tabular}{lcrrr} \hline
   &&\multicolumn{3}{c}{Energy $E_0$ [GeV]}\\ \cline{3-5}
   & $Z$       & $3\cdot10^8$ & $5\cdot10^8$ & $10^9$ \\ \hline
   protons     &1             &(7)  &(11)   &(17)    \\
   helium      &2-5           &(19)  &(28)   &(37)    \\
   CNO         &6-14          &(10)+2&(12)+1 &(13)    \\
   heavy       &15-27         & 22   & 10    &  4     \\
   ultra heavy &28-92         & 40   & 38    & 29     \\
 \hline
 \end{tabular}
\end{table}

Using the HiRes data for the fit would result in a slightly lighter mass
composition. But due to the relatively large errors for the energy range
$5-10\cdot10^8$~GeV a fit seems not to be meaningful.

The fraction of protons obtained for the \adhoc component is plotted in
\fref{pfract} as function of energy (filled points). Between $10^7$ and
$10^9$~GeV the primary cosmic-ray flux is not dominated by protons.  Between
$6\cdot10^6$~GeV and $5\cdot10^8$~GeV the fraction amounts to less than 10\%,
between $10^7$~GeV and $2\cdot10^8$~GeV its value is even below 5\%.  One has
to bear in mind this indication when proton-air cross-sections are derived from
air shower measurements.
 
The low inelastic cross-sections and higher values for the elasticity assumed
in model~3a have implications on the mass composition derived from \Xmax\
measurements at energies above $10^7$~GeV as pointed out above.  With the low
cross-sections also intermediate and heavy elements are important in this
energy region.  In the energy region between $10^7$ and $10^8$~GeV a relatively
heavy composition has been found also by Alvarez-Mu\~niz \etal (2002b), viz.
consisting of 85\% Fe, 10\% CNO, 4\% He, and 1\% protons.  Similar values are
calculated with the \modell\ at $4\cdot10^7$~GeV: 86\% heavy and ultra-heavy
($9\le Z\le92$), 9\% CNO ($6\le Z\le8$), 4\% He ($2\le Z\le5$), and $<1$\%
protons.  A key issue of the present investigations is that cosmic rays above
$10^8$~GeV, i.e. the \adhoc component, contain a significant contribution of
particles heavier than protons.  A mass composition heavier than protons only
in this region is also obtained by Erlykin and Wolfendale (2002).

\section{Mean logarithmic mass}\label{lnasect}

Many scientists characterize the mass composition of high-energy cosmic rays by
the mean logarithmic mass. It is defined as
\begin{equation}\label{lna} 
 \lna\equiv \sum_i r_i \ln A_i \quad, 
\end{equation} 
with the relative fraction $r_i$ of nuclei with mass $A_i$.  Knowing the
average depth of the shower maximum for protons $X_{max}^p$ and iron nuclei
$X_{max}^{Fe}$ from simulations, the mean logarithmic mass can be derived in
the superposition model from the measured values $X_{max}^{meas}$ using 
\begin{equation}\label{xmaxfun}
\lna=\frac{X_{max}^{meas}-X_{max}^{p}}{X_{max}^{Fe}-X_{max}^{p}} 
   \cdot \ln A_{Fe}\quad .
\end{equation}

\begin{figure}[bt] \centering
 \epsfig{file=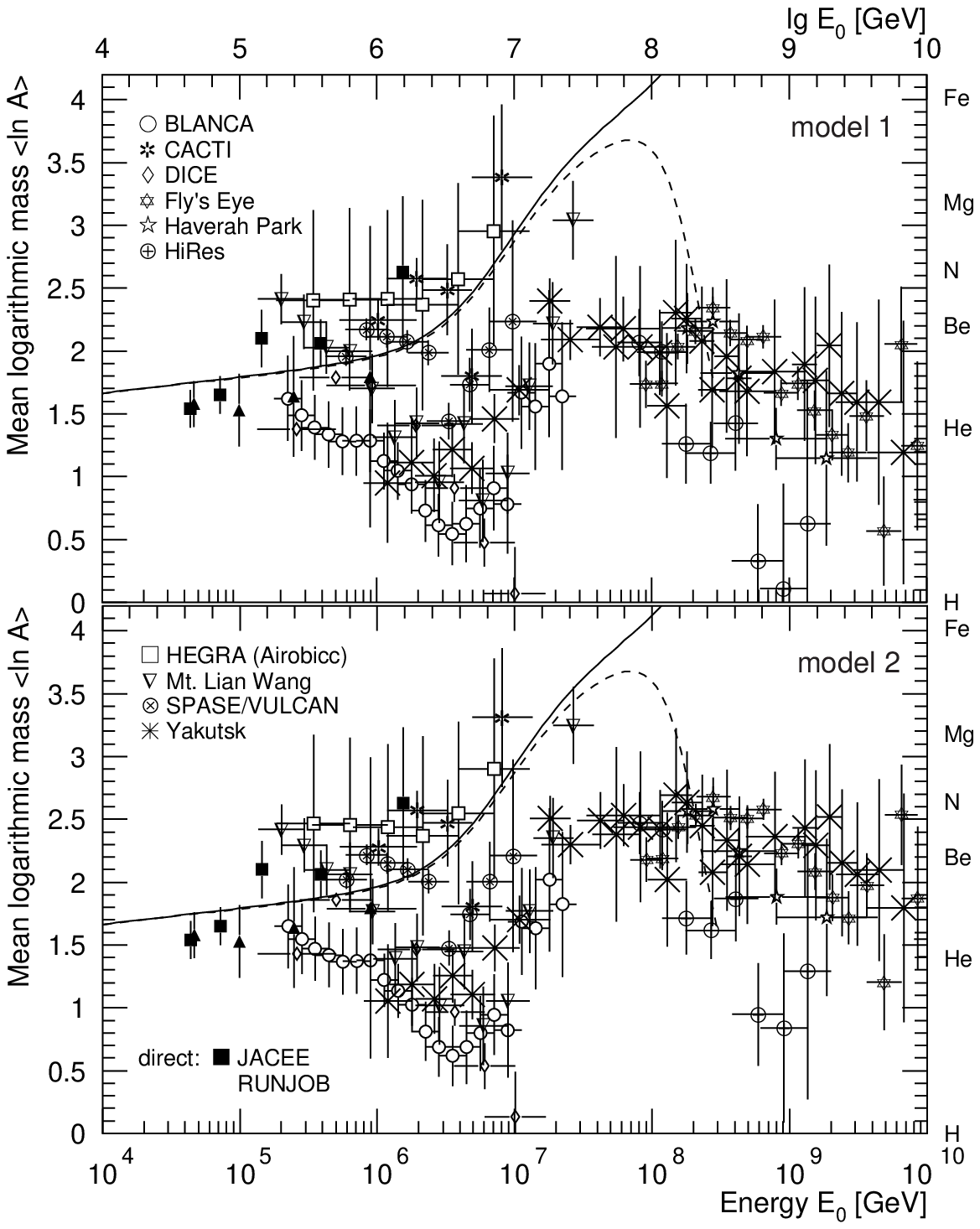,width=0.8\textwidth} 
 \caption{Mean logarithmic mass $\lna$ of primary cosmic rays versus their 
	  energy calculated from experimental \Xmax\ values from air shower
	  observations (see \fref{xmax}) with the results of
	  CORSIKA/QGSJET simulations for two different sets of inelastic
	  hadronic cross-sections, model~1 (top) and model~2 (bottom).  Results
	  of direct measurements from JACEE (Shibata 1999) and RUNJOB
	  (Apanasenko \etal 2001) are shown as well.  The lines represent
	  calculations according to the \modell: galactic component only
	  (\full), plus \adhoc component of solely protons (\dashed).}
 \label{masse1}         
\end{figure} 

\begin{figure}[bt] \centering
 \epsfig{file=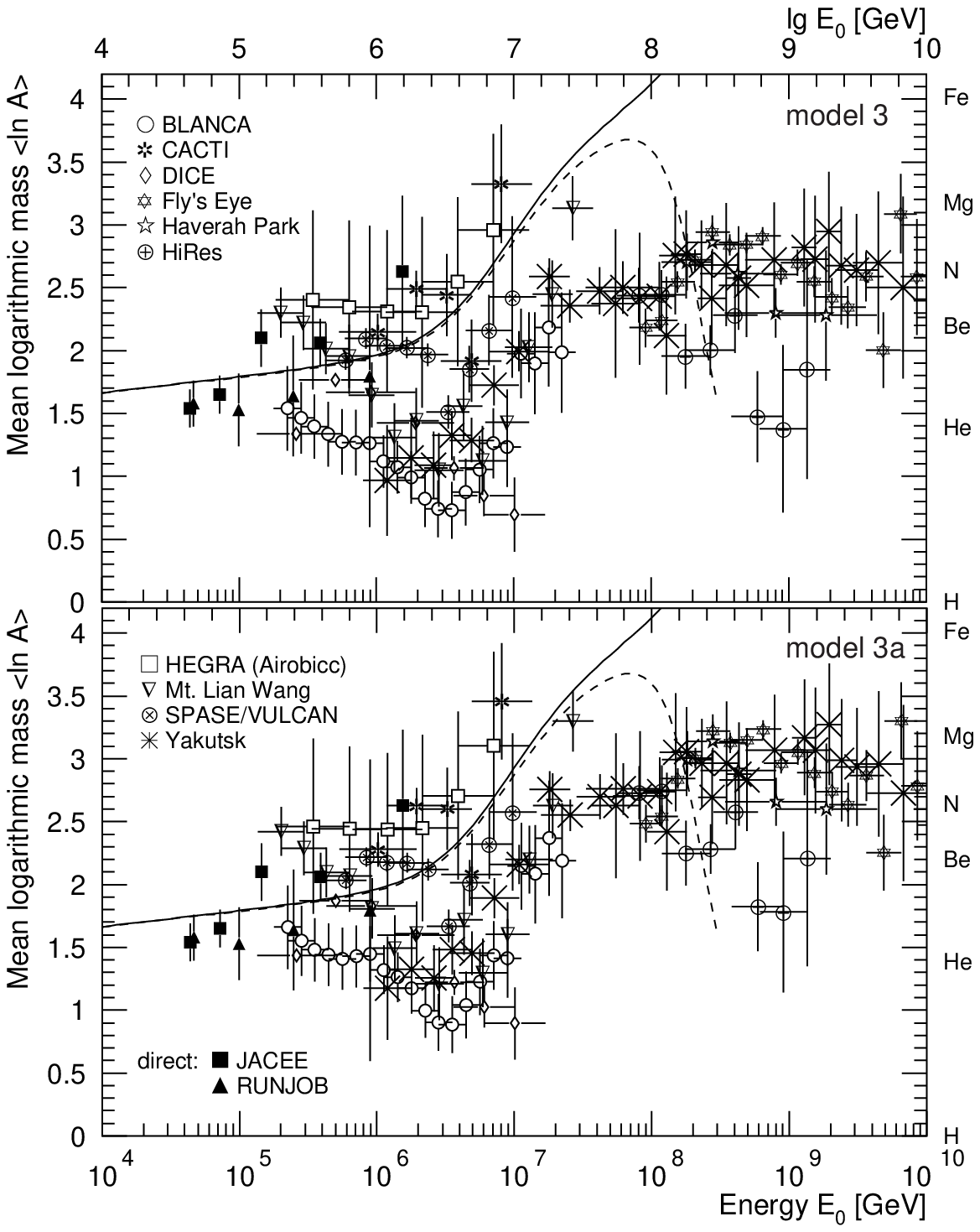,width=0.8\textwidth} 
 \caption{Mean logarithmic mass $\lna$ of primary cosmic rays versus their 
	  energy calculated from experimental \Xmax\ values from air shower
	  observations (see \fref{xmax}) with the results of CORSIKA/QGSJET
	  simulations using models~3 (top) and 3a (bottom).
	  See also caption of \fref{masse1}.}
 \label{masse3}         
\end{figure}

\begin{figure}[btp] \centering
 \epsfig{file=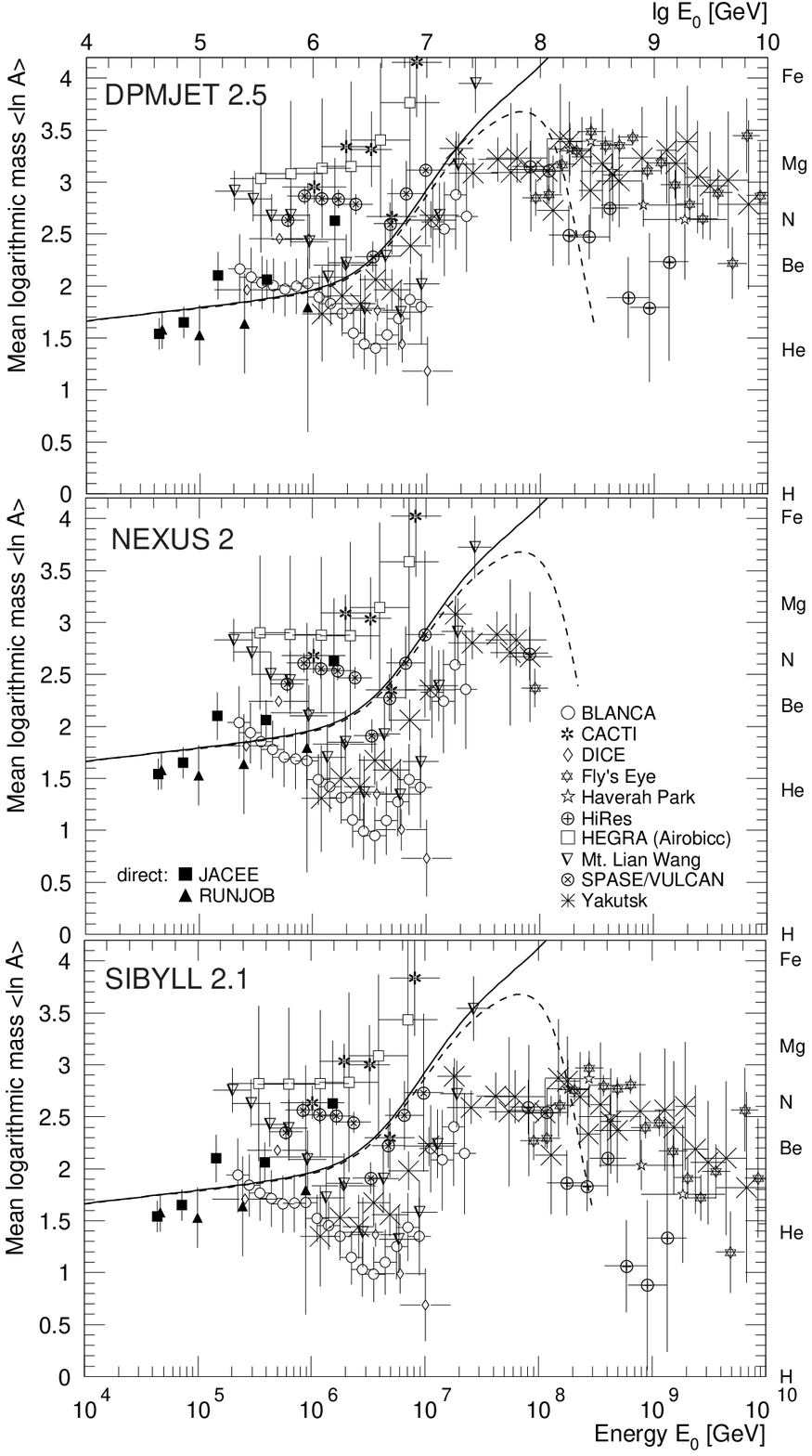,height=0.89\textheight} 
 \caption{Mean logarithmic mass $\lna$ of primary cosmic rays versus their 
	  energy calculated from experimental \Xmax\ values from air shower
	  observations (see \fref{xmax}) with the results of CORSIKA
	  simulations for three different hadronic interaction models as
	  indicated, see \fref{xmaxmod}.  See also caption of
	  \fref{masse1}.}
 \label{massecors}         
\end{figure} 

The corresponding $\lna$ values for the variations of QGSJET~01, obtained from
the data presented in \fref{xmax}, are plotted versus the primary energy in
\fref{masse1} for models~1 and 2 as well as for models~3 and 3a in
\fref{masse3}.  The average $\lna$ increases as the cross-sections decrease
from model~1 to model~3.  For the original QGSJET the results of many
experiments exhibit a (strong) decrease of $\lna$ up to about $4\cdot10^6$~GeV
and an increase above this energy. The energy of this dip in the $\lna$ values
coincides with the energy of the \knie\ in the all-particle energy spectrum.
The dip becomes less striking with lower inelastic cross-sections and higher
values for the elasticity.  For model~3a only a modest dip can be recognized.
At 4~PeV the average values increase from $\overline{\lna}=1.2$ for model~1 to
$\overline{\lna}=1.6$ for model~3a.  Around $10^8$~GeV the average logarithmic
mass compared with model~1 is about 
$\Delta\lna\approx0.5$ larger for model~3 and
$\Delta\lna\approx0.7$ larger for model~3a.
These examples illustrate how strong the interpretation of air shower
measurements depends on model parameters like the inelastic cross-sections or
elasticities used.  At Tevatron energies the cross-sections vary within the
error range given by the experiments and at $10^8$~GeV the proton-air
cross-sections of models 1 and 3 differ only by about 10\%, but the general
trend of the emerging $\lna$ distributions proves to be significantly
different.

At this point the circle closes. If we assume in the energy region from $10^7$
to $10^8$~GeV a small proton fraction only, the cross-sections have to be
corrected and lowered to the region of values for model~3. In turn a heavier
composition is obtained for model~3 as has been demonstrated in
\fref{masse3}. Thus, at least qualitatively the arguments are consistent.

Results from the balloon experiments JACEE (Shibata 1999) and RUNJOB
(Apanasenko \etal 2001) are presented in figures~\ref{masse1} to
\ref{massecors} (filled points) for comparison.  No hint for a decreasing mean
logarithmic mass is indicated by these measurements.  The solid lines shown in
the figures are predictions according to the \modell\ for the galactic
component and the dashed lines include an \adhoc component of protons only.
Both, the results of the balloon experiments shown as well as the values
calculated with the \modell\ seem to support model~3a.

In addition to the modifications of QGSJET, it would be interesting to look at
the mean mass which is deduced from the measurements applying other models.
\Xmax\ values for the models DPMJET, NEXUS, and SIBYLL have been presented in
\fref{xmaxmod}.  Taking these predictions the experimental results lead to mean
logarithmic masses shown in \fref{massecors}.  At 4~PeV the average mass values
range from $\overline{\lna}=1.7$ for SIBYLL and NEXUS to $\overline{\lna}=2.0$
for DPMJET.  For NEXUS the authors of the model guarantee validity up to
$10^8$~GeV, as mentioned above, for this reason $\lna$ is shown up to this
energy only.  As can be inferred from \fref{xmaxmod}, very similar depths are
obtained for the models NEXUS and SIBYLL, both are hardly discernable from
model~2.  As a consequence, the mean logarithmic masses derived are very much
alike.  On the other hand, the depths predicted by DPMJET are very similar to
the results of model~3a and for both models a heavier mass composition is
obtained.  For the program MOCCA with its internal interaction model the
showers penetrate extremely deep into the atmosphere and a relative heavy mass
composition is obtained, not compatible with direct measurements (H\"orandel
2003).

At \knie\ energies the average experimental $\lna$ values vary from 1.2 when
the data are interpreted with the original QGSJET (model~1) to 2.0 for a DPMJET
interpretation. One has to admit that the model ambiguities result in an
uncertainty $\Delta\lna\approx 0.8$.  At this energy the scatter in the
measured average depth of the shower maximum, as presented in \fref{xmax},
yields a rms value of the individual experimental results in the order of
$\mbox{rms}_{\lna}\approx0.6$.  This value is compatible with
$\mbox{rms}_{\lna}\approx0.6$ obtained for experiments measuring particle
distributions at ground level, as discussed elsewhere (H\"orandel 2003).

Also other authors have studied the effects of different interaction models on
$\lna$.  The systematic influence of the models HDPM, QGSJET, SIBYLL, and VENUS
on the results of the BLANCA experiment is discussed by Fowler \etal (2001).  At
\knie\ energies a maximum offset $\Delta\lna\approx0.8$ between QGSJET and
SIBYLL as well as HDPM is obtained.  The investigations of Wibig (2001) yield a
maximum difference $\Delta\lna\approx0.9$ between VENUS and SIBYLL.  Comparing
results from QGSJET~98 and SIBYLL~1.6 an uncertainty of $\Delta\lna\approx0.3$
is found by Swordy \etal (2002) at \knie\ energies.  Investigations of several
hadronic observables by the KASCADE group yield model ambiguities of
$\Delta\lna\approx0.4$ around the \knie\ (H\"orandel \etal 1998).  These
estimates from the literature seem to be well compatible with the ambiguities
determined above.

As already mentioned in the introduction, the mean logarithmic mass as obtained
with experiments investigating particle distributions at ground level is
compatible with the results of the \modell. But these results disagree with
$\lna$ values obtained from measurements of the longitudinal shower development
interpreted with QGSJET~01.  This incompatibility can be seen in \fref{masse1},
as well as for the models NEXUS~2 and SIBYLL~2.1 presented in \fref{massecors}.
For several experiments the mean logarithmic mass decreases as function of
energy up to \knie\ energies, a tendency not supported by the extrapolation of
the direct measurements. This effect is strongest for QGSJET~01.  

Using lower cross-sections and larger values for the elasticity in the model
QGSJET the $\lna$ values obtained are comparable with the results of
experiments investigating particle distributions. In other words, the
disagreement between the two groups of experiments can be reduced, if model~3a
is taken.  With the altered inelastic cross-sections and the larger elasticity
consistency can be achieved between the predictions of the \modell\ and the
mass composition derived from observed \Xmax\ values.  Similar values of $\lna$
as for model~3a are also obtained using DPMJET~2.5 to interpret the data.  The
main conclusion of the present investigation is that relatively deep
penetrating showers with \Xmax\ values similar to the ones obtained with
model~3a or DPMJET~2.5 seem to yield  most consistent $\lna$ values.  On the
other hand, investigations of secondary particles produced in air showers by
Milke \etal (2001), based on CORSIKA simulations with the low energy model
GHEISHA, reveal that for a given number of muons DPMJET~2.5 transports more
hadronic energy to the observation level as compared with the measurements of
the KASCADE experiment. Hence, presently model~3a is the preferred model to
describe air shower measurements most consistently.

\section{Number of electrons and muons} \label{nenmusect}

\begin{figure}[b] \centering
 \epsfig{file=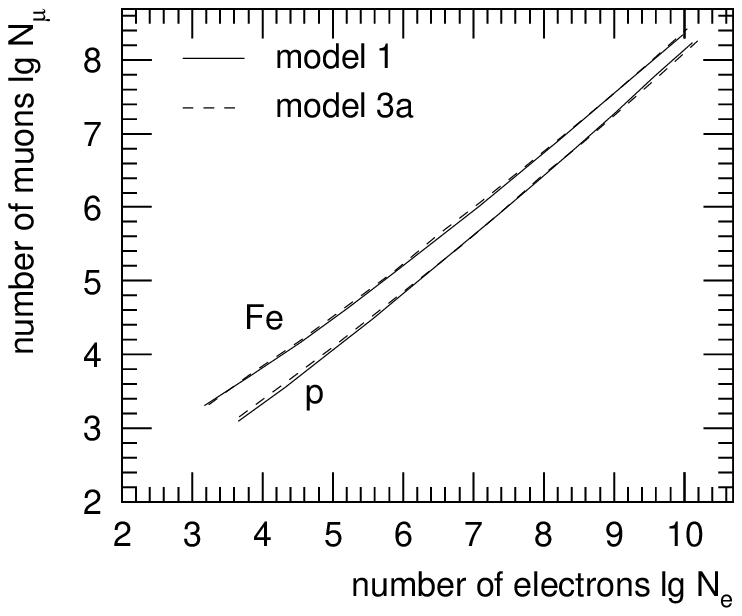,width=0.49\textwidth}
 \epsfig{file=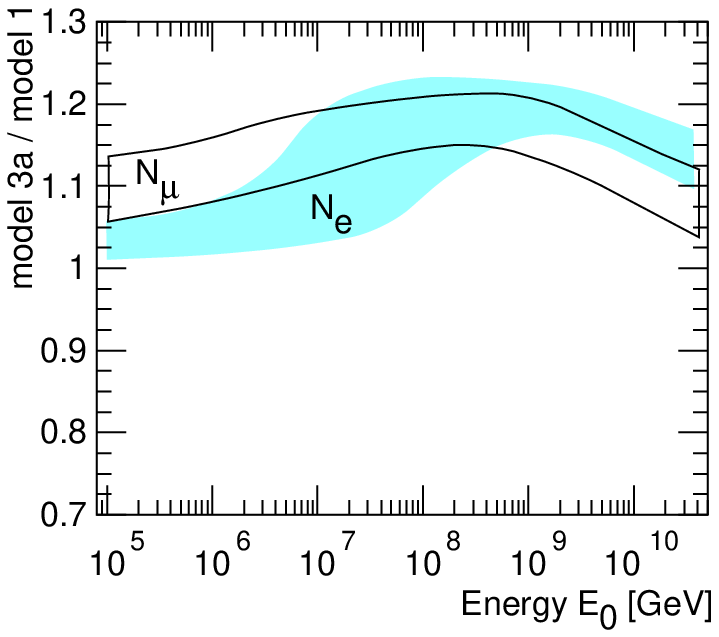,width=0.45\textwidth}
 \caption{Left-hand panel: Average number of muons ($E_\mu>100$~MeV) versus
	  number of electrons ($E_e>0.25$~MeV) for model~1 and 3a. The range
	  pictured corresponds to primary energies from $10^5$ to
	  $3\cdot10^{10}$~GeV.  Right-hand panel: Number of muons and
	  electrons as predicted by model~3a relative to the numbers calculated
	  with model~1 as function of primary energy.  The range for the ratio
	  $N_\mu^{3a}/N_\mu^1$ for the extrema protons and iron nuclei is
	  illustrated by solid lines and the range for the ratio
	  $N_e^{3a}/N_e^1$ is represented as shaded area.}
 \label{kartoffel}         
\end{figure}

The modified cross-sections and elasticity values influence not only the
average depth of the shower maximum but also other air shower observables. The
implications on shower sizes at ground level might be strong and, therefore,
are investigated in the following. Many air shower arrays use the correlation
between the number of electrons $N_e$ and the number of muons $N_\mu$ to
determine the cosmic-ray mass composition.  For showers with primary energies
between $10^5$ and $10^{10}$~GeV the average number of muons with energies
above 100~MeV is plotted versus the average number of electrons with energies
above 0.25~MeV in \fref{kartoffel} (left-hand side). Correlations from
CORSIKA/QGSJET simulations for proton and iron induced showers are shown, using
the original QGSJET~01 and model~3a. The first impression is, no significant
differences between the two models can be found and, consequently, no
significant changes in the mass composition derived are expected.

A closer look, when investigating the relative changes for the numbers obtained
reveals small deviations.  The number of electrons and muons obtained with
model~3a normalized to the results of the original QGSJET model are shown on
the right-hand side of the figure.  At energies of the \knie\ (4~PeV) the
number of electrons increases by about 5\% when using model~3a for both
primaries, protons and iron nuclei. The increase in the number of muons amounts
to about 15\%. This outcome may qualitatively explain the need to accommodate
the energy scales of many air shower experiments with respect to direct
measurements as outlined in the introduction.  It has been found that at \knie\
energies air shower experiments overestimate the energy on average by about
3\%. This can be understood, if the interaction models predict too small
electron and muon numbers, since these quantities are mostly used by
experiments to estimate the primary energy.  In fact, the change in electron
and muon numbers from models~1 to 3a corresponds to an energy uncertainty of
about 5\% and 15\%, respectively.  These numbers depend on the threshold for
electrons as well as muons and differ from experiment to experiment. A
quantitative investigation for each experiment would require detailed detector
simulations.  However, the numbers indicate that the observed energy
overestimation of air shower experiments relative to direct measurements may be
explained by too high inelastic cross-sections and too low values for the
elasticity.

At higher energies around $10^8$~GeV the number of both, electrons and muons,
generated by model 3a is increased by about 15\% to 20\% as compared with the
original QGSJET (model~1).  The changes in $N_e$ and $N_\mu$ correspond to
energy shifts in the order of 10\% to 20\%.  In this energy region the air
shower experiments overestimate the primary energy on average by about 10\%
according to the above mentioned investigations. This value is comparable with
the systematic offset between models~1 and 3a. Again, the exact numbers depend
on individual detection thresholds for each experiment.  These findings are an
independent hint for a slower logarithmic rise of the inelastic cross-sections
as well as slightly increased elasticities and, consequently, deeper
penetrating cascades as presumed so far.

It may be summarized that differences between the models shift the data points
parallel to the lines on the left-hand side of \fref{kartoffel} (i.e. in
"energy direction") and not perpendicular (i.e. in "mass direction"), hence
only a change in energy and no significant change in the mass composition
derived from the data are expected.  More detailed investigations are necessary
to study the implications of the modifications on observables, performing
detector simulations for individual experiments, but this is beyond the scope
of the present article.  In addition, to interpret experimental data not only
high-energy interaction models as discussed presently are important, but also
models to describe interactions below 100~GeV, as for example GHEISHA, FLUKA,
or UrQMD, all available in CORSIKA.  For a final conclusion, the complex
interplay between low and high-energy models as well as their influence on
observables measured in ground arrays has to be studied.

\section{Summary and conclusion}
The impact of reduced inelastic cross-sections and increased values for the
elasticity on the development of extensive air showers has been studied with
the simulation program CORSIKA.  Within the high-energy hadronic interaction
model QGSJET the logarithmic increase of the inelastic cross-sections as
function of energy has been lowered and the elasticity has been increased.  The
average depth of the shower maximum was calculated for primary protons and iron
nuclei and compared with experimental values. The mean logarithmic mass $\lna$
was derived from the experimental data for three assumptions for the increase
of the inelastic cross-sections and two hypotheses for the elasticity.

The disagreement in the observed $\lna$ values between experiments measuring
particle distributions at ground level and experiments measuring the average
depth of the shower maximum could be reduced. The data exhibit still a
scattering in the order of $\Delta\lna\approx0.5$ but the general trend of the
increase as function of energy is now similar for both classes.
A reduced and only modest
increase of the cross-sections as function of energy had to be applied.  Best
agreement is obtained for model~3a with an increase of the total inelastic
proton-proton cross-section to 72~mb at $10^9$~GeV and an additional increase
of the elasticity of about 10\% to 15\%.  In turn, the increase of $\lna$ with
energy according to the \modell\ becomes compatible with the two classes of
experiments.  As a result, model~3a allows a consistent description of the
extrapolations of individual element spectra as obtained by direct measurements
and the all-particle energy spectrum as well as the cosmic-ray mass composition
obtained by most air shower experiments.

\ack
The author would like to thank 
S.~Ostapchenko for giving insight into details of the model QGSJET and
D.~Heck for support related to the CORSIKA simulation program.
It is a great pleasure to acknowledge fruitful discussions with 
R.~Engel, J.~Engler, and S.~Ostapchenko 
as well as the encouraging support by H.~Bl\"umer.

\References
 \item[] Abe F \etal 1990  \PR D {\bf 41} 2330
 \item[] \dash       1994a \PR D {\bf 50} 5518
 \item[] \dash       1994b \PR D {\bf 50} 5550
 \item[] Abu-Zayyad T \etal 2000 \PRL {\bf 84} 4276
 \item[] \dash              2001 \ApJ {\bf 557} 686
 \item[] Aglietta M \etal 1997 \ICRC{25th}{Durban} {\bf 6} 37
 \item[] Alexopoulos T \etal 1993 \PR D {\bf 48} 984
 \item[] \dash               1998 \PL B {\bf 435} 453
 \item[] Alvarez-Mu\~niz J \etal 2002a \PR D {\bf 66} 033011
 \item[] \dash                   2002b \PR D {\bf 66} 123004
 \item[] Amos N A \etal 1990 \PL B {\bf 243} 158
 \item[] \dash          1992 \PRL {\bf 68} 2433	 
 \item[] Anchordoqui L A \etal 1999 \PR D {\bf 59} 094003
 \item[] Anderson B \etal 1991 {\it Z. Phys.} C {\bf 50} 405
 \item[] Antoni T \etal 1999 \jpg {\bf 25} 2161
 \item[] Apanasenko A V \etal 2001 \ICRC{27th}{Hamburg} {\bf 5} 1622
 \item[] Arqueros F \etal 2000 \AA {\bf 359} 682
 \item[] Avila C \etal 1999 \PL B {\bf 445} 419
 \item[] Baltrusaitis R M \etal 1984 \PRL {\bf 52} 1380
 \item[] Bird D J \etal 1994 \ApJ {\bf 424} 491
 \item[] Block M M \etal 1992 \PR D {\bf 45} 839
 \item[] \dash           2000 \PR D {\bf 62} 077501
 \item[] Bossard G \etal 2001 \PR D {\bf 63} 054030
 \item[] B\"uttner C \etal 2001 \ICRC{27th}{Hamburg} {\bf 1} 153
 \item[] Capdevielle \etal 1992 {\it Report KfK 4998}, 
         Kernforschungszentrum Karlsruhe
 \item[] Capdevielle J N and Attallah R 1995 \jpg {\bf 21} 121	 
 \item[] Cha M \etal 2001 \ICRC{27th}{Hamburg} {\bf 1} 132
 \item[] Dickinson J E \etal 1999 \ICRC{26th}{Salt Lake City} {\bf 3} 136
 \item[] Drescher H J \etal 2001 {\it Phys. Rep.} {\bf 350} 93
 \item[] Dyakonov M N \etal 1990 \ICRC{21st}{Adelaide} {\bf 9} 252
 \item[] \dash              1993 \ICRC{23rd}{Calgary} {\bf 4} 303
 \item[] Engel R \etal 1998 \PR D {\bf 58} 014019	 
 \item[] \dash         1999 \ICRC{26th}{Salt Lake City} {\bf 1} 415
 \item[] Erlykin A D and Wolfendale A W 2002 \ApP {\bf 18} 151
 \item[] Fletcher R S \etal 1994 \PR D {\bf 50} 5710
 \item[] Frichter G M \etal 1997 \PR D {\bf 56} 3135
 \item[] Fowler G N \etal 1987 \PR D {\bf 35} 870
 \item[] Fowler J W \etal 2001 \ApP {\bf 15} 49
 \item[] Gaisser T K and Halzen F 1987 \PRL {\bf 54} 1754
 \item[] Gaisser T K \etal 1987 \PR D {\bf 36} 1350     
 \item[] \dash             1993 \PR D {\bf 47} 1919     
 \item[] Glauber R J and Matthiae G 1970 \NP B {\bf 21} 135
 \item[] Hagiwara K \etal (Particle Data Group) 2002 \PR D {\bf 66} 010001
 \item[] Hara T \etal 1983 \PRL {\bf 50} 2058	 
 \item[] Heck D \etal 1998 {\it Report FZKA 6019}, Forschungszentrum Karlsruhe;
	 and http://www-ik3.fzk.de/$\sim$heck/corsika/ 
 \item[] Hillas A M 1997 \NP B (Proc. Suppl.) {\bf 52B} 29
 \item[] Honda M \etal 1993 \PRL {\bf 70} 525	 
 \item[] H\"orandel J R \etal 1998 {\it Proc. 16th European Cosmic Ray
         Symposium} (Alcala de Henares) 579
 \item[] H\"orandel J R 2003 \ApP {\bf 19} 193
 \item[] Huang J \etal 2003 \ApP {\bf 18} 637
 \item[] Keilhauer B \etal 2003 {\it Auger Technical Note} GAP-2003-009
 \item[] Kalmykov N N \etal 1995 \ICRC{24th}{Rome} {\bf 1} 123
 \item[] \dash              1997 \NP B (Proc. Suppl.) {\bf 52B} 17
 \item[] Knapp J \etal 2003 \ApP {\bf 19} 77
 \item[] Knurenko S P \etal 1999 \ICRC{26th}{Salt Lake City} {\bf 1} 372
 \item[] \dash              2001 \ICRC{27th}{Hamburg} {\bf 1} 177
 \item[] Kopeliovich B Z \etal 1989 \PR D {\bf 39} 769
 \item[] Landau L D 1969 {\it Men of Physics: L D Landau-II}, edited by
         D ter Haar, Pergamon, New York, p 131
 \item[] Mielke H H \etal 1994 \jpg {\bf 20} 637	 
 \item[] Milke J \etal 2001 \ICRC{27th}{Hamburg} {\bf 1} 241
 \item[] Pajares C \etal 2000 \ApP {\bf 12} 291
 \item[] Paling A \etal 1997 \ICRC{25th}{Durban} {\bf 5} 253
 \item[] Pryke C L 2001 \ApP {\bf 14} 319	 
 \item[] Ranft J 1995 \PR D {\bf 51} 64
 \item[] \dash   1999 {\it Preprint} hep-ph/9911213
 \item[] Sciutto S J 1998 {\it Auger technical note} GAP-98-032;
         AIRES users guide and reference manual, available at
         http://www.fisica.unlp.edu.ar/auger/aires/eg\_Aires.html
 \item[] Shibata T 1999 \NP B (Proc. Suppl.) {\bf 75A} 22	 
 \item[] Swordy S P and Kieda D B 2000 \ApP {\bf 13} 137
 \item[] Swordy S P \etal 2002 \ApP {\bf 18} 129
 \item[] Watson A A 2000 {\it Phys. Rep.} {\bf 333 - 334} 309	 
 \item[] Werner K 1993 {\it Phys. Rep.} {\bf 232} 87
 \item[] Wibig T 1997 \PR D {\bf 56} 4350
 \item[] \dash   1999 \jpg {\bf 25} 557 
 \item[] \dash   2001 \jpg {\bf 27} 1633
 \item[] Wibig T and Sobczy\'nska D 1998 \jpg {\bf 24} 2037
 \item[] Yodh G B \etal 1972 \PRL {\bf 28} 1005
 \item[] \dash          1973 \PR D {\bf 8} 3233
 \item[] \dash          1983 \PR D {\bf 27} 1183
\endrefs

\end{document}